\documentclass[%
 reprint,
superscriptaddress,
 amsmath,amssymb,
 aps,
pra,
]{revtex4-2}

\usepackage{dsfont}
\usepackage{svg}
\usepackage{graphicx}
\usepackage{dcolumn}
\usepackage{physics}
\usepackage{amsmath}
\usepackage{bbold}
\usepackage{float}
\usepackage{tcolorbox}

\usepackage{hyperref}
\hypersetup{
    colorlinks=true,
	citecolor=blue,
    linkcolor=red,
	urlcolor = blue
}

\usepackage{bm}
\usepackage[mathlines]{lineno}%

\newcommand{\Trs}{\mathrm{Tr}}
\newcommand{\MF}{\mathrm{MF}}
\definecolor{newtext}{RGB}{140,0,140}

\begin{document}

\title{Second-order mean force Gibbs state of fermionic open systems}

\author{David J. Strachan}
\email{david.strachan@bristol.ac.uk}
\affiliation{\textit{H. H. Wills Physics Laboratory, University of Bristol, Bristol BS8 1TL, United Kingdom}}

\date{\today}   
\begin{abstract}

Thermodynamics typically assumes that a system in contact with a thermal reservoir relaxes to the Gibbs state, or to the grand canonical Gibbs state when it also exchanges particles. However, this only holds when the coupling is very weak. Beyond this, the expected equilibrium state is the mean force Gibbs (MFG) state, obtained by tracing the reservoir out of the thermal state of the combined system. Despite its relevance, general analytic results for the MFG state have so far been obtained almost exclusively for bosonic reservoirs. Here we derive the grand canonical MFG state to second order in the coupling for a generic fermionic system coupled to a thermal reservoir at arbitrary chemical potential, using the concept of the mode reduced state in order for the fermionic MFG state to be well defined. The correction to the bare Gibbs state separates into particle and hole channels weighted by the empty and occupied reservoir states, related through the Kubo-Martin-Schwinger condition. We apply the result to the resonant level model, two modes coupled to a shared reservoir, and the single impurity Anderson model. For the two modes, the reservoir mediates coherence between modes that do not interact directly, something the bare Gibbs state cannot capture. We verify our results against numerically exact calculations, including real-time dynamics for the interacting model, and find agreement between the static prediction and the dynamical relaxation.

\end{abstract}
\maketitle

Thermalisation is a cornerstone of thermodynamics: a system placed in contact with a much larger environment at temperature $T$ and chemical potential $\mu$ generically relaxes to a state characterised by these two parameters alone. In classical statistical mechanics this is the grand canonical Boltzmann distribution, quantum mechanically it is the grand canonical Gibbs state
\begin{equation}
\tau_{S} = \frac{e^{-(H_{S}-\mu N_{S})/k_{B}T}}{Z_{S}},
\label{eq:intro_1}
\end{equation}
with $Z_{S}\equiv \Trs\big(e^{-(H_{S}-\mu N_{S})/k_{B}T}\big)$ the grand partition function. This state is determined entirely by the system Hamiltonian $H_S$, the system particle number $N_S$, and the temperature and chemical potential of the environment~\footnote{When the environment exchanges only energy with the system, $N_S$ is fixed and Eq.~\eqref{eq:intro_1} reduces to the canonical Gibbs state $e^{-H_S/k_BT}/\Trs(e^{-H_S/k_BT})$.}. Equation~\eqref{eq:intro_1} underpins much of quantum thermodynamics, it fixes the free state in thermodynamic resource theories~\cite{Horodecki2013,Brandao2015} and is the reference equilibrium against which the performance of quantum batteries~\cite{RevModPhys.96.031001,Campaioli2018} and nanoscale heat engines~\cite{Uusnakki2026,Bhattacharjee2021} are assessed.

That the equilibrium state should depend only on $H_S$ and $N_S$ is, however, an assumption. It rests on the interaction energy between system and environment being negligible compared with the system's own energy scales, so that the interaction does not contribute to the form of the equilibrium state. This is an accurate assumption for macroscopic systems but fails for microscopic ones, precisely the regime quantum thermodynamics is concerned with. For example, in quantum dots and molecular junctions the system-environment coupling is routinely comparable to both the level spacing and the temperature~\cite{Josefsson2018,Burkard2023,CuevasScheer2017,ThossEvers2018}. This motivates the field of strong coupling thermodynamics~\cite{RevModPhys.92.041002,10.1116/5.0073853,Seifert2016,PhysRevX.7.011008,Rivas2020,Miller2018,PhysRevE.95.062101}, which retains the system--environment interaction explicitly in thermodynamic quantities. A central object of the field is the mean force Gibbs (MFG) state~\cite{RevModPhys.92.041002,10.1116/5.0073853}. Suppose system and environment have jointly equilibrated, so that their combined state is the \emph{total} grand canonical Gibbs state
$\tau_{\mathrm{tot}}=e^{-(H-\mu N)/k_{B}T}/\Trs(e^{-(H-\mu N)/k_{B}T})$ with $H=H_{S}+H_{R}+\lambda V$ where $\lambda V$ is the system-reservoir interaction with coupling strength $\lambda$ and $N=N_S+N_R$ the total particle number. This is the appropriate description whenever the combined object is itself macroscopic, either because it has equilibrated with a weakly coupled ``super-bath''~\cite{RevModPhys.83.771,Kosloff2019} at temperature $T$ and chemical potential $\mu$, or by appeal to the eigenstate thermalisation hypothesis~\cite{PhysRevA.43.2046,PhysRevE.50.888}. The reduced state of the system is then the (grand canonical) MFG state
\begin{equation}
\tau_{\MF} = \Trs_{R}\big(\tau_{\mathrm{tot}}\big).
\label{eq:intro_2}
\end{equation}
Note that in the following, Gibbs and MFG state refer to the grand canonical versions unless stated otherwise. In the limit of weak coupling, this reduces to the Gibbs state of Eq.~\eqref{eq:intro_1}. Unlike $\tau_S$, the MFG state need not commute with $H_S$: it can retain coherences in the bare energy eigenbasis~\cite{PhysRevLett.127.250601,Guarnieri2018,Purkayastha2020,LewisWilliamson2026}, a feature with no classical counterpart and one that is absent from the steady state predicted by standard secular master equations. This is a particularly attractive feature, as coherence is often seen as a resource~\cite{RevModPhys.89.041003}, for example to improve the efficiency of heat engines~\cite{PhysRevLett.122.110601}.

Despite its compact definition, Eq.~\eqref{eq:intro_2} is hard to evaluate and analytic progress has been limited to particular models or to asymptotic coupling regimes. Of particular note is the work of Cresser and Anders~\cite{PhysRevLett.127.250601}, who obtained closed-form expressions for an arbitrary system coupled to a bosonic reservoir in both the weak and ultrastrong coupling limits, subsequently extended to anharmonic environments~\cite{KumarGhosh2024} for the ultrastrong limit, multiple channels~\cite{TimofeevTrushechkin2022} and complemented by effective-Hamiltonian~\cite{AntoSztrikacs2023} and numerically exact~\cite{Chiu2022,Cerisola2024} treatments. However, these works are limited to the bosonic case. For fermions, even the definition of the MFG state given in Eq.~\eqref{eq:intro_2} is ambiguous as the partial trace operation is defined for a Hilbert space with a tensor product structure, which fermions do not possess due to their canonical anti-commutation relation (CAR) algebra. This necessitates the use of the fermionic partial trace, defined via the mode reduced state~\cite{Amosov_2016,PhysRevA.87.022338}. The usual coupling operator is also different to the standard bosonic case, where for a canonical fermionic open system the interaction transfers a particle between system and reservoir
\begin{equation}
V = X^{\dagger}E + E^{\dagger}X ,
\label{eq:intro_3}
\end{equation}
where $X$ is an odd-parity system operator, typically an annihilation operator, and $E$ is a linear combination of reservoir annihilation operators. This differs from $X\otimes B$ with Hermitian factors considered in~\cite{PhysRevLett.127.250601}.

In this work we close this gap. Our main result, Eq.~\eqref{eq:result_2}, is a closed-form expression for the MFG state of a generic fermionic system coupled through Eq.~\eqref{eq:intro_3} to a number-conserving fermionic reservoir up to second order in the coupling. We use the fermionic partial trace as defined in~\cite{Amosov_2016,PhysRevA.87.022338} ensuring its validity. The Kubo-Martin-Schwinger condition fixes all coefficient functions in terms of a single reservoir function. We derive both the low and high temperature  limits where the Gibbs state is recovered at high temperature and coherences in the bare energy eigenbasis persist at low temperature. We also derive the corresponding Hamiltonian of mean force. Exchange coupling is also a legitimate choice for bosonic reservoirs and Eq.~\eqref{eq:result_2} covers both statistics, although its main content is the fermionic case.

We verify Eq.~\eqref{eq:result_2} against numerically exact calculations, showing that at weak and intermediate coupling the steady state agrees with the MFG state predicted by our formula. There is no general proof that the MFG state is the long-time state of an open system~\cite{10.1116/5.0073853}, and the existing numerical and analytical evidence mostly concerns bosonic environments~\cite{Chiu2022,Cerisola2024,LewisWilliamson2026,KumarAthulyaGhosh2025}. So for all the examples we consider, we compare our static formula with the steady state obtained from real-time dynamics. For quadratic Hamiltonians, which include the first two examples we consider, both the exact dynamics and the exact MFG state follow directly from the single-particle correlation matrix~\cite{Peschel2003} (see also Sec.~\ref{sec:SM-static-C}). For interacting systems no such result is available.  For the single impurity Anderson model we obtain the steady state from real-time dynamics, using a matrix product state (MPS) description of the reservoir based on the thermofield chain mapping~\cite{deVegaBanuls2015,Nusseler2020,Strachan2024,PhysRevB.104.014303}. For all three examples we find agreement with our static prediction, providing evidence that the MFG state is indeed the relevant equilibrium state for fermionic open systems.

The paper is organised as follows. Section~\ref{sec:Setup} introduces the setup and the fermionic partial trace. Section~\ref{sec:main-result} presents the main result and its general properties. Section~\ref{sec:numerics} outlines the numerical approach we use to benchmark our result, which is applied in Section~\ref{sec:examples} to the resonant level model, two modes coupled to a shared reservoir, the single impurity Anderson model. The Supplemental Material contains the derivations for Eq.~\eqref{eq:result_2}, the high- and low-temperature limits, the Hamiltonian of mean force, the examples and additional details on the non-interacting case.

\section{Setup} \label{sec:Setup}
We consider the canonical open-system setting
\begin{equation}
H = H_{S}+H_{R}+\lambda V,
\label{eq:setup_1}
\end{equation}
where $H_{S}$ is the system Hamiltonian, $H_R$ the reservoir Hamiltonian and $\lambda V$ their interaction. We use the standard continuum description for a Gaussian reservoir
\begin{equation} \label{eq:setup_2}
H_{R} = \int \! d\varepsilon \;\varepsilon\, a^{\dagger}_{\varepsilon}a_{\varepsilon},
\end{equation}
\begin{equation}
V = \!\int \! d\varepsilon\sqrt{J(\varepsilon)}\Big(a^{\dagger}_{\varepsilon}X+X^{\dagger}a_{\varepsilon}\Big),
\label{eq:setup_3}
\end{equation}
where $a^{\dagger}_{\varepsilon},a_{\varepsilon}$ create and annihilate a reservoir excitation at frequency $\varepsilon$, satisfying $\{a_{\varepsilon},a^{\dagger}_{\varepsilon'}\}=\delta(\varepsilon-\varepsilon')$ for a fermionic reservoir and $[a_{\varepsilon},a^{\dagger}_{\varepsilon'}]=\delta(\varepsilon-\varepsilon')$ for a bosonic one. We use natural units $k_B=\hbar=1$ throughout and write $\beta\equiv1/T$ for the inverse temperature. Here $\lambda$ is the coupling strength, which for the normalisation $\int d \varepsilon J(\varepsilon)=1$ used below carries units of energy. $X$ is a system operator and $J(\varepsilon)$ is the reservoir spectral density. Note that $X$ need not be Hermitian and in general will not be, rather it is usually a fermionic annihilation operator such that $V$ describes hoppings between the system and the environment at each frequency $\varepsilon$ with strength $\lambda\sqrt{J(\varepsilon)}$. We take the coupling to conserve the total particle number $N=N_S+N_R$, where $N_R=\int d\varepsilon\,a^{\dagger}_{\varepsilon}a_{\varepsilon}$ and $N_S$ is the system number operator, with $[H_S,N_S]=0$. Since $[N_R,a_\varepsilon]=-a_\varepsilon$, this requires $[N_S,X]=-X$, i.e.\ $X$ removes a single particle from the system, as is the case when $X$ is a fermionic annihilation operator. Then $[H,N]=0$, and the global equilibrium state entering Eq.~\eqref{eq:intro_2} is the grand canonical Gibbs state $\tau_{\rm tot}\propto e^{-\beta(H-\mu N)}$.  Its generator can be written as
\begin{equation}
\begin{gathered}
H-\mu N = K_S+K_R+\lambda V,\\
K_S\equiv H_S-\mu N_S,\qquad K_R\equiv\int\! d\varepsilon\,(\varepsilon-\mu)\, a^{\dagger}_{\varepsilon}a_{\varepsilon},
\end{gathered}
\label{eq:setup_4}
\end{equation}
so that the system Gibbs state of Eq.~\eqref{eq:intro_1} is $\tau_S=e^{-\beta K_S}/Z_S$. 

As mentioned in the introduction, for fermions the partial trace in Eq.~\eqref{eq:intro_2} needs care. This can be dealt with via the definition of the reduced state. For a state $\rho$ of the system and reservoir, the reduced state $\rho_S \equiv\textrm{Tr}_{R}(\rho)$ is the unique operator on the system Fock space $\mathcal{F}_{S}$ such that 
\begin{equation}  \label{eq:setup_5}
\textrm{Tr}_{S}(\rho_{S}O_{S}) = \textrm{Tr}(\rho O_{S}) \quad \forall O_{S} \in \mathcal{A}_{S},
\end{equation}
where $\mathcal{A}_{S}$ is the algebra generated by the system operators $s_{i}^{\dag}s_{i},s_{i}s_{i}^{\dag},s_{i},s_{i}^{\dag}$ for $i=1,...,L_S$ where $L_S$ is the number of system modes and $\{\hat{s}^{\dag}_{i},\hat{s}_{j}\} = \delta_{ij}$. This defines what is meant by $\textrm{Tr}_{R}$ in Eq.~\eqref{eq:intro_2}, which coincides with the partial trace for bosons due to the underlying tensor product structure, but for fermions this is not always the case~\cite{PhysRevA.87.022338,Amosov_2016}. To define states in a Fock space, a specific mode ordering must be chosen, for example in the full Fock space $\mathcal{F}$ states can be defined as 
\begin{equation} \label{eq:setup_6}
\ket{\boldsymbol{n}_{S},\boldsymbol{m}_{R}}= (s_{1}^{\dag})^{n_1}...(s_{L_{S}}^{\dag})^{n_{L_{S}}}(a_{1}^{\dag})^{m_{1}}...(a_{L_R}^{\dag})^{m_{L_{R}}}\ket{\textrm{vac}},
\end{equation}
where $L_{S},L_{R}$ are the number of system and reservoir modes respectively and we have put the system modes first, where for notational simplicity we have used a finite, discrete set of reservoir modes. In~\cite{Amosov_2016}, it is shown that if the modes being traced out form a contiguous set in the given mode ordering and the global state $(\tau_{\textrm{tot}}$ in our case) is parity symmetric which is guaranteed from superselection rules~\cite{Amosov_2016,PhysRevD.1.3267,PhysRev.88.101}, then the reduced state $\rho_{S}$ coincides with the operator obtained from the conventional partial trace under a Jordan Wigner transform (See Sec.~\ref{app:fermionic partial trace} for more details). We thus fix the mode ordering as Eq.~\eqref{eq:setup_6} from the start.

\section{Second order MFG state and Hamiltonian of Mean Force} \label{sec:main-result}

Here we present the main result, a closed form expression for the mean force Gibbs state for a total fermionic Hamiltonian given by Eqs.~(\ref{eq:setup_1}-\ref{eq:setup_3}) up to second order in $\lambda$, with an error of $O(\lambda^{4})$. We decompose the coupling operator in the eigenoperator basis of $K_S$,
\begin{equation}
X=\sum_{\omega}X_{\omega},
\qquad
X_{\omega}=\!\!\sum_{n,m:\,\varepsilon_{n}-\varepsilon_{m}=\omega}\!\! \Pi_{n}X\Pi_{m},
\label{eq:result_1}
\end{equation}
where $[K_{S},X_{\omega}]=\omega X_{\omega}$ and $\Pi_n$ projects onto the (possibly degenerate) eigenspace of $K_S$ with eigenvalue $\varepsilon_n$. Projecting onto distinct Bohr frequencies rather than distinct transitions allows for degenerate energy gaps. Because every matrix element of $X$ lowers $N_S$ by one, each $\omega$ is a Bohr frequency of $H_S$ shifted by $\mu$, and the operators $X_\omega$ coincide with those obtained by decomposing $X$ with respect to $H_S$.
Combining Eq.~\eqref{eq:result_1} with the Kubo expansion of $e^{-\beta(H-\mu N)}$~\cite{Thingna2012,Geva2000,MoriMiyashita2008,PhysRevLett.127.250601} and performing the reservoir trace (See Sec.~\ref{sec:SM-derivation} for details) gives the MFG state
\begin{align} \label{eq:result_2}
\tau_{\MF}^{(2)}
&=\tau_{S}+\lambda^{2}\big(\Delta_{\rm diag}+\Delta_{\rm off}\big),
\end{align}
where
\begin{align}
\Delta_{\rm diag} &=\sum_{\omega}
 \tau_{S}\Big(X_{\omega}^{\dagger}X_{\omega}-\Trs_{S}\big[\tau_{S}X_{\omega}^{\dagger}X_{\omega}\big]\Big)D_{1}(\omega)
 \nonumber\\
&\hspace{1.1em}
 +\tau_{S}\Big(X_{\omega}X_{\omega}^{\dagger}-\Trs_{S}\big[\tau_{S}X_{\omega}X_{\omega}^{\dagger}\big]\Big)D_{2}(\omega),
\label{eq:result_3}
\\[4pt]
\Delta_{\rm off} &=\sum_{\omega\neq\omega'}
    (\omega'-\omega)^{-1}\Big[
   \big(X_{\omega'} X_{\omega}^{\dagger}\tau_{S}+\tau_{S}X_{\omega} X_{\omega'}^{\dagger}\big) A_2(\omega)
   \nonumber\\
    &\hspace{4.6em}- \big(X_{\omega'}^{\dagger} X_{\omega}\tau_{S}+\tau_{S}X_{\omega}^{\dagger} X_{\omega'}\big)A_1(\omega)
    \Big].
\label{eq:result_4}
\end{align}
The four coefficient functions are imaginary-time transforms of the two reservoir correlation functions
\begin{equation}
G_{1}(\beta,-i\tau)=\Trs_R\!\big[\tau_R E(-i\tau)E^{\dagger}\big],\label{eq:result_5}
\end{equation}
\begin{equation}
G_{2}(\beta,-i\tau)=\Trs_R\!\big[\tau_R E^{\dagger}(-i\tau)E\big],
\label{eq:result_6}
\end{equation}
where $E = \int d\varepsilon\sqrt{J(\varepsilon)}a_{\varepsilon}$, and the imaginary-time evolution is generated by $K_R$, $E(-i\tau)=e^{\tau {K_R}}Ee^{-\tau K_R}$.
Namely,
\begin{equation}
A_{1}(\omega)=\int_{0}^{\beta}\!d\tau\,e^{-\tau\omega}G_{1}(\beta,-i\tau),\label{eq:result_7}
\end{equation}
\begin{equation}
A_{2}(\omega)=\int_{0}^{\beta}\!d\tau\,e^{\tau\omega}G_{2}(\beta,-i\tau),
\label{eq:result_8}
\end{equation}
\begin{align}
D_{1}(\omega)&=\beta A_1(\omega)+A_1'(\omega),
&
D_{2}(\omega)&=\beta A_2(\omega)-A_2'(\omega).
\label{eq:result_9}
\end{align}

Closed forms for $A_{1,2}$ and $D_{1,2}$ in terms of $J(\varepsilon)$ and the Fermi (or Bose) function are given in Sec.~\ref{sec:SM-coefficients}, Eqs.~\eqref{eq:SM_derivation_47}--\eqref{eq:SM_derivation_50}. Two correlation functions appear, one for each
term of the exchange coupling. $G_1$ virtually propagates a particle into the reservoir, and $G_2$ virtually propagates a hole~\footnote{The term virtual here means the propagation is in imaginary time.}. They are not independent, however. Because the reservoir is thermal, the imaginary-time Kubo--Martin--Schwinger condition $G_1(\beta,-i\tau)=G_2\big(\beta,-i(\beta-\tau)\big)$ holds, and (see Sec.~\ref{sec:SM-KMS})
\begin{equation}
A_{1}(\omega)=e^{-\beta\omega}A_{2}(\omega),
\qquad
D_{1}(\omega)=e^{-\beta\omega}A_{2}'(\omega).
\label{eq:result_10}
\end{equation}
All four coefficients are therefore fixed by $A_2$ and its derivative. Since every $\omega$ is a Bohr frequency of $H_S$ shifted by $\mu$, the denominators $\omega'-\omega$ of $\Delta_{\rm off}$ and those inside $A_{1,2}$ and $D_{1,2}$ do not depend on $\mu$; the chemical potential enters only through $\tau_S$ and through the reservoir occupation. An equivalent form, in which the coefficient functions remain bounded as $T\to0$ and which makes both the low-temperature limit and the comparison with the bosonic result of Ref.~\cite{PhysRevLett.127.250601} direct, is given in Sec.~\ref{sec:SM-lowT}. It is also possible to write $\tau_{\textrm{MF}}^{(2)}$ in exponential form, in terms of a Hamiltonian of mean force. Writing $\tau_{\textrm{MF}}\propto e^{-\beta (H_{\textrm{MF}}-\mu N_S)}$, we have (see Sec.~\ref{sec:HMF})
\begin{align}
  H_{\MF} &= H_S+\lambda^2K^{(2)}+\mathcal{O}(\lambda^4),\nonumber\\
  K^{(2)} &= -\frac{1}{\beta}\sum_\omega\Big[D_1(\omega)\,X^\dagger_\omega X_\omega
      +D_2(\omega)\,X_\omega X^\dagger_\omega\Big] \nonumber \\
  &-\sum_{\omega\neq\omega'}\bigg[
      \frac{A_2(\omega')-A_2(\omega)}{e^{\beta\omega'}-e^{\beta\omega}}\,X^\dagger_\omega X_{\omega'} \nonumber \\
     &+\frac{A_1(\omega')-A_1(\omega)}{e^{-\beta\omega'}-e^{-\beta\omega}}\,X_\omega X^\dagger_{\omega'}\bigg],
  \label{eq:result_11}
\end{align}
where $H_{\MF}=K_{\MF}+\mu N_S=H_S+\lambda^2K^{(2)}$ because $K^{(2)}$ conserves
$N_S$. Note that this form is more restrictive than Eq.~\eqref{eq:result_2}. Gibbs state calculated from Eq.~\eqref{eq:result_11} reproduces Eq.~\eqref{eq:result_2} only when the mean force corrections to the populations are small compared with the bare populations themselves. As $T\to0$ the bare populations of excited states vanish exponentially while their corrections remain finite, so the exponential form is restricted to sufficiently high temperatures, whereas Eq.~\eqref{eq:result_2} is not (see Sec.~\ref{sec:HMF}).

Both $X_{\omega}^{\dagger}X_{\omega}$ and $X_{\omega}X_{\omega}^{\dagger}$ commute with $H_S$, so $\Delta_{\rm diag}$ is block diagonal with respect to the eigenspaces of $H_S$, and diagonal if $H_S$ is non-degenerate~\footnote{When $H_S$ has degeneracies, $\Delta_{\rm diag}$ may carry coherences within a degenerate block, though these are basis-conventional rather than physical, since any basis of a degenerate eigenspace is equally an energy eigenbasis}. The off-diagonal sum is not,
\begin{align}
[\tau_{\MF}^{(2)}, H_S] &= \lambda^2 \!\!\sum_{\omega\neq\omega'}\!
\Big\{ A_2(\omega)\big[\tau_S X_\omega X_{\omega'}^{\dagger} - X_{\omega'}X_\omega^{\dagger}\,\tau_S\big] \nonumber \\
&\hspace{3.2em}+A_1(\omega)\big[\tau_S X_\omega^{\dagger} X_{\omega'} - X_{\omega'}^{\dagger} X_\omega\,\tau_S\big] \Big\},
\label{eq:result_12}
\end{align}
which is generically non-zero. As in the bosonic case~\cite{PhysRevLett.127.250601}, the equilibrium state retains coherences across energy sectors in the eigenbasis of $H_S$. The high- and low-temperature limits are derived in Secs.~\ref{sec:SM-highT} and~\ref{sec:SM-lowT} respectively. At high temperature the correction vanishes, $\tau_{\MF}^{(2)} = \tau_{S}+\mathcal{O}(\lambda^{2}\beta^{2})$, faster than the $\mathcal{O}(\beta)$ deviation of $\tau_S$ itself from the maximally mixed state, and as $\mathcal{O}(\lambda^{2}\beta^{3})$ when $X$ is linear in system annihilation operators, as in all three examples below. At low temperature the coherences persist.

\section{Numerical methods} \label{sec:numerics}

To validate Eq.~\eqref{eq:result_2}, we compare to numerics. When the system Hamiltonian is quadratic in mode operators, $H_{S}=\sum_{ij}\boldsymbol{h}_{ij}s^{\dag}_{i}s_{j}$, the total Hamiltonian $H=\sum_{ij}\boldsymbol{H}_{ij}d^{\dag}_{i}d_{j}$ is also quadratic, where $\{d_{i}\}$ denotes both the system and reservoir modes in the ordering of Eq.~\eqref{eq:setup_6}. For Gaussian initial states, both the real-time dynamics and the MFG state can then be obtained exactly from the single particle correlation matrix $\boldsymbol{C}_{ij}(t) = \textrm{Tr}[\rho(t)d_{j}^{\dag}d_{i}]$, which evolves unitarily as (see Sec.~\ref{sec:sm-noninteracting})
\begin{equation} \label{eq:numerics_1}
\boldsymbol{C}(t) = e^{-i\boldsymbol{H}t}\boldsymbol{C}(0)e^{i\boldsymbol{H}t},
\end{equation}
where $\rho(t)$ is the combined state of the system and reservoir. For the MFG state, we write the generator of Eq.~\eqref{eq:setup_4} as $H-\mu N=\sum_{ij}\boldsymbol{K}_{ij}\,d_i^{\dag}d_j$, so that the correlation matrix of the global Gibbs state is $\boldsymbol{C}=f(\boldsymbol{K})$, where $f(w)=(e^{\beta w}+1)^{-1}$. As the MFG state is Gaussian, it is determined by the system block $\boldsymbol{C}_S=[f(\boldsymbol{K})]_{i,j\in S}$ alone~\cite{Peschel2003,PhysRevB.69.075111} (see  also Sec.~\ref{sec:SM-static-C}),
\begin{equation} \label{eq:numerics_2}
\tau_{\MF}=\frac{\exp\!\big(-\sum_{ij}\boldsymbol{M}_{ij}\,s_i^{\dag}s_j\big)}{\Trs_S\exp\!\big(-\sum_{ij}\boldsymbol{M}_{ij}\,s_i^{\dag}s_j\big)},
\qquad
\boldsymbol{M}=\ln\!\big(\boldsymbol{C}_S^{-1}-\mathbb{1}\big),
\end{equation}
so that $\beta^{-1}\boldsymbol{M}$ is the single particle matrix of $H_{\MF}-\mu N_S$, and occupations and coherences can be read off directly as $\langle s_j^{\dag}s_i\rangle_{\MF}=(\boldsymbol{C}_S)_{ij}$. For interacting systems there is no exact solution in general, and a numerical method must be used. For this we use the thermofield chain mapping~\cite{deVegaBanuls2015,Nusseler2020,PhysRevB.104.014303,Strachan2024}, computing the real-time dynamics until convergence from an initial product state $\rho(0)=\rho_0\tau_{R}$. The thermal reservoir is purified with an auxiliary copy, and a Bogoliubov rotation of the doubled reservoir gives two independent reservoirs with spectral densities $J(\varepsilon)f(\varepsilon-\mu)$ and $J(\varepsilon)[1-f(\varepsilon-\mu)]$, which are initially filled and empty respectively. Each reservoir is mapped onto a semi-infinite tight-binding chain using orthogonal polynomials~\cite{Chin2010}, with the system coupled only to the first site of each chain. We represent the system and both chains as an MPS and compute the dynamics using the time-dependent variational principle~\cite{Haegeman2016} as implemented in ITensor~\cite{Fishman_2022}. For more details on the approach see~\cite{Strachan2024,PhysRevB.104.014303}. We use chains of $O(100)$ sites, a time step $\delta t=0.1/D$, a maximum bond dimension of $400$ and a truncation cutoff of $10^{-9}$, all of which we have checked for convergence. With the system placed first in the mode ordering, its reduced state coincides with the fermionic reduced state (Sec.~\ref{app:fermionic partial trace}). For the quadratic models we evolve Eq.~\eqref{eq:numerics_1} in the same chain geometry~\cite{Strachan2024}. To evaluate Eq.~\eqref{eq:numerics_2} we instead apply the orthogonal polynomial mapping to the physical spectral density $J(\varepsilon)$. We find $\boldsymbol{C}_S$ converged to $10^{-12}$ when the chain is truncated at $200$ sites.

\section{Examples} \label{sec:examples}

We now consider some examples. For the two quadratic models we compute the exact MFG state via Eq.\eqref{eq:numerics_2}, and have confirmed it agrees with the long-time state of the exact correlation-matrix dynamics as described by Eq.~\eqref{eq:numerics_1} for both the chain mapping discretisation discussed in Sec.~\ref{sec:numerics} and linear discretisation~\cite{PhysRevB.92.155126}. We therefore refer to both as the exact result. For the interacting SIAM we compare Eq.\eqref{eq:result_2} to the converged real-time dynamics using the thermofield approach as described in Sec.~\ref{sec:numerics}. For all cases we choose a semi-elliptical spectral density $J(\varepsilon) = \frac{2}{\pi D}\sqrt{1-(\varepsilon/D)^{2}}$, although it must be emphasised that this choice is relatively arbitrary and not the focus of the discussion. $D$ is the half-bandwidth, which we will use as our unit of energy. This definition fixes the normalisation as $\int_{-D}^{D}J(\varepsilon)d\varepsilon = 1$. In each example below we compare three states: the bare Gibbs state $\tau_{S}$, our prediction $\tau_{\MF}^{(2)}$, and the (numerically) exact result.
\begin{figure}[t!]
  \centering
  \includegraphics[width=0.49\textwidth]{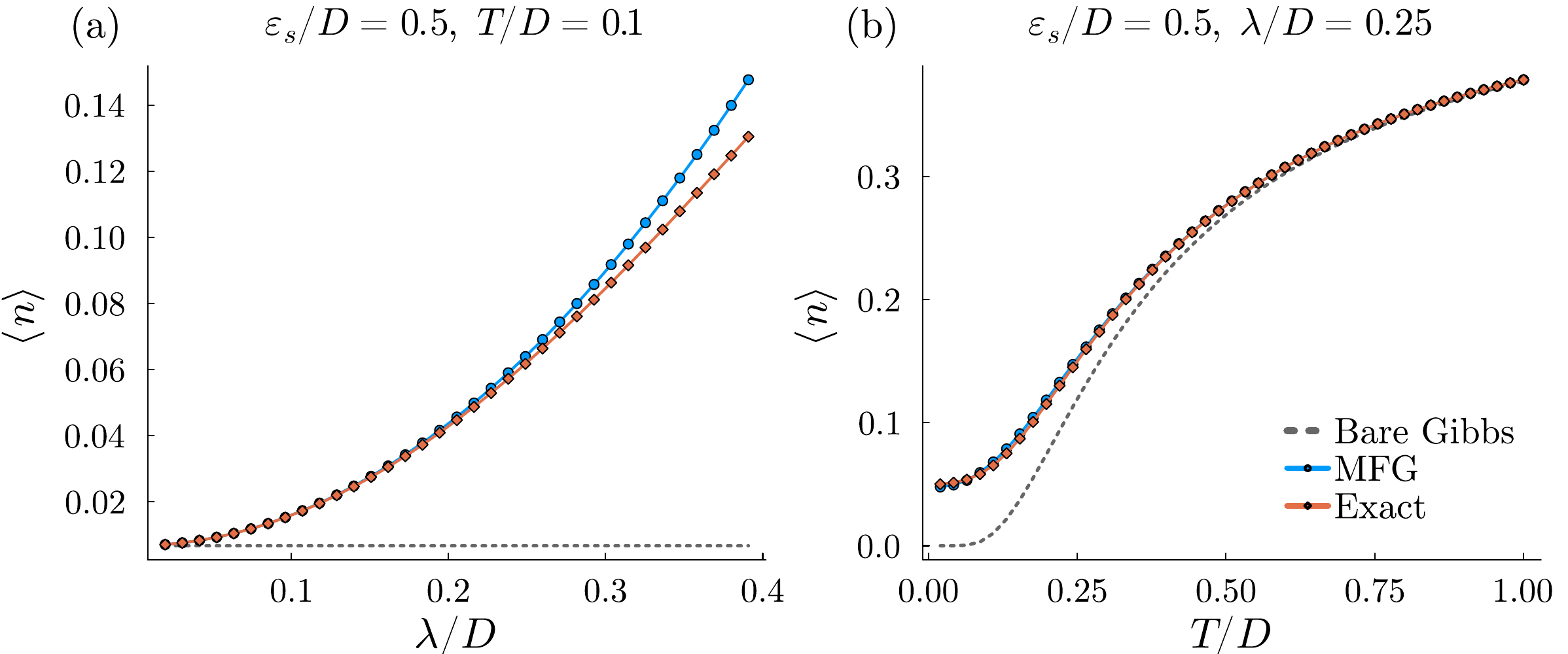}
  \caption{Steady state occupation of the resonant level model $H_S = \varepsilon_s s^{\dag}s$ across (a) coupling at fixed temperature and (b) temperature at fixed coupling, with the bare Gibbs, MFG and exact predictions shown. The energy of the level is fixed at $\varepsilon_s = 0.5D$ with $\mu=0$. }
  \label{fig:RLM}
\end{figure}

\textit{Resonant level model} - We begin with the resonant level model (RLM): a single fermionic mode with $H_S=\varepsilon_s n$, $n \equiv s^{\dag}s$, coupled to the reservoir through $X=s$. Here $X$ is already an eigenoperator with $\omega_0=\mu-\varepsilon_s$, so $\Delta_{\rm off}$ is empty and Eq.~\eqref{eq:result_2} gives (see Sec.~\ref{sec:SM-RLM})
\begin{equation} \label{eq:examples_1}
\langle n\rangle^{(2)}_{\textrm{MF}} = f_s+\lambda^{2}f_s\big(1-f_s\big)\big[D_{1}(\omega_0)-D_{2}(\omega_0)\big],
\end{equation}
where $f_s\equiv f(\varepsilon_s-\mu)$ and $f$ is the Fermi Dirac distribution. This admits a clear physical interpretation and relates to the standard Green's function treatment of the RLM~\cite{Ochoa2016,Dou2018}. Defining $\langle n\rangle_\MF^{(2)}=f_s+\delta\langle n\rangle$, the deviation from the bare Gibbs value can be decomposed as (see Sec.~\ref{sec:SM-RLM})
\begin{equation}
\delta\langle n\rangle = \frac{\partial f_s}{\partial\varepsilon_s}\,\Lambda(\varepsilon_s)
\;-\;\lambda^{2}\,\mathcal{P}\!\!\int\! d\varepsilon\, J(\varepsilon)\,\frac{f_s-f(\varepsilon-\mu)}{(\varepsilon-\varepsilon_{s})^{2}},\label{eq:examples_2}
\end{equation}
\begin{equation}
\Lambda(\varepsilon_s)\equiv \lambda^{2}\,\mathcal{P}\!\!\int\! d\varepsilon\,\frac{J(\varepsilon)}{\varepsilon_{s}-\varepsilon}.\label{eq:examples_3}
\end{equation}
The first term describes a level shift by $\Lambda$ due to the hybridisation with the reservoir, which the occupation responds to with the Fermi-function susceptibility. The second term represents the broadening of the level, sampling $f$ away from $\varepsilon_s$.

At low temperatures the coefficient functions diverge, leading to numerical precision issues, so we instead use the bounded form derived in Sec.~\ref{sec:SM-lowT} for the points at lowest $T$, although the numerical differences between forms are negligible above $T\sim 0.05D$ for all models considered. Figure~\ref{fig:RLM}(a) shows $\langle n\rangle$ at fixed $T/D=0.1$ against $\lambda$. The bare Gibbs state predicts the level to be essentially empty at this temperature. As the coupling grows, hybridisation with the reservoir shifts and broadens the level, giving a finite occupation that increases with $\lambda$. Our perturbative MFG prediction captures this, matching the exact result at low to intermediate coupling and departing from it at strong coupling as expected. Figure~\ref{fig:RLM}(b) shows the same quantity against temperature at fixed $\lambda/D=0.25$: the occupation remains finite as $T\to0$, and both curves converge to the Gibbs state at high temperature.

\textit{Two independent modes coupled to a shared reservoir} - We now consider a model that allows for coherence in the bare system energy eigenbasis. Specifically, we consider two modes that do not interact directly but share a reservoir. An interesting feature of this model is that any coherences that arise must be mediated by the reservoir rather than directly from interactions between the two modes. This is the fermionic analogue of~\cite{LewisWilliamson2026} in which they study reservoir mediated coherences in the mean force Gibbs state for a multi-qubit spin boson problem. The Hamiltonian is given by
\begin{equation}
H_S=\varepsilon_1 n_1+\varepsilon_2 n_2, \label{eq:examples_4}
\end{equation}
with coupling operator $X=s_1+s_2$. Each mode is an eigenoperator, $X_{\omega_1}=s_1$, $X_{\omega_2}=s_2$ with frequencies $\omega_1=\mu-\varepsilon_1$, $\omega_2=\mu-\varepsilon_2,$ and $X =X_{\omega_1}+X_{\omega_2}$. As in~\cite{LewisWilliamson2026}, there exists a decoherence-free state when $\varepsilon_1 =\varepsilon_2$. The antisymmetric combination $d_{-}=(s_1-s_2)/\sqrt{2}$ does not appear in $X=s_1+s_2=\sqrt{2}\,d_{+}$ and for $\varepsilon_{1}=\varepsilon_{2}$ there is no hopping between $d_+$ and $d_{-}$ in $H_{S}$, so $d_{-}$ would be entirely decoupled from the reservoir, known as a dark state~\cite{Buca2019,PhysRevLett.81.2594}. In this case the MFG need not be the state the system relaxes to, as this will depend on the initial occupation of the dark state. To avoid this, we consider $\varepsilon_1 \neq\varepsilon_2$.

\begin{figure}[t!]
  \centering
  \includegraphics[width=0.49\textwidth]{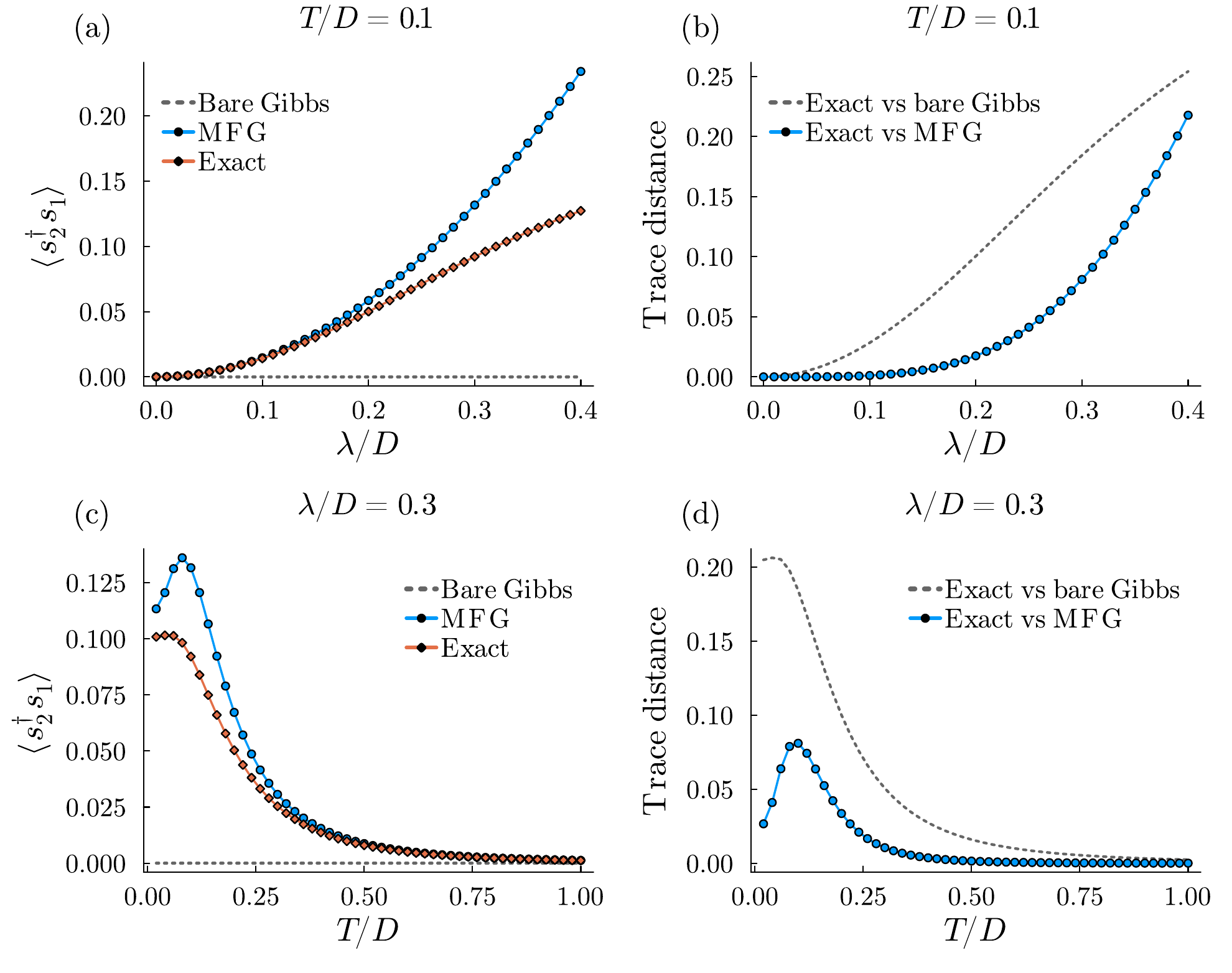}
  \caption{Comparison of the steady state as predicted by Eq.~\eqref{eq:result_2}, the exact result and the bare Gibbs state of the two mode model $H_S = \varepsilon_1 n_1+\varepsilon_2 n_2$. (a) Coherence against coupling at fixed temperature $T/D=0.1$. (b) Trace distance between predictions against coupling at $T/D=0.1$. (c) Coherence against temperature at fixed coupling $\lambda/D=0.3$ (d) Trace distance between predictions against temperature at fixed coupling $\lambda/D=0.3$. The energies of the two levels are fixed at $\varepsilon_1,\varepsilon_2 = 0.3D,0.4D$, with $\mu=0$.}
  \label{fig:two-mode}
\end{figure}

At second order in the coupling, each mode's population is blind to the other's presence and the occupations are given by Eq.~\eqref{eq:examples_1} applied to each mode respectively (see Sec.~\ref{sec:SM-two-mode}). The modes are not totally blind to each other however, as they would be in the bare Gibbs state, they acquire a coherence, given by 
\begin{align}
\big\langle s_2^\dagger s_1\big\rangle_{\tau^{(2)}_{\mathrm{MF}}}
  &=\lambda^2\,\frac{\Phi(\varepsilon_1)-\Phi(\varepsilon_2)}{\varepsilon_1-\varepsilon_2},
  \label{eq:examples_5}
\end{align}
where $\Phi(\varepsilon)$ is defined in Sec.~\ref{sec:SM-two-mode} along with additional details on $\tau_{\textrm{MF}}^{(2)}$. This is reflected through the Hamiltonian of mean force. Evaluating Eq.~\eqref{eq:result_11} gives a reservoir-mediated hopping term,
\begin{equation}
  H_{\mathrm{MF}}=\sum_{i=1,2}\tilde\varepsilon_i\,n_i
  +g\,\big(s_1^\dagger s_2+s_2^\dagger s_1\big),
  \label{eq:examples_6}
\end{equation}
\begin{equation}
g=\lambda^2\,\frac{\Phi(\varepsilon_1)-\Phi(\varepsilon_2)}{f_1-f_2},\label{eq:examples_7}
\end{equation}
where $\tilde{\varepsilon}_{i}$ are defined in the same way as the renormalised energy of the Hamiltonian of mean force for the RLM, and we have omitted an overall constant in $H_{\textrm{MF}}$ (see Sec.~\ref{sec:SM-two-mode}). Note that Eqs.~\eqref{eq:examples_6}--\eqref{eq:examples_7} are only valid at sufficiently high temperature (see Sec.~\ref{sec:HMF}), but the coherence, Eq.~\eqref{eq:examples_5}, holds more generally. Figures~\ref{fig:two-mode}(a), (c) show Eq.~\eqref{eq:examples_5} and the exact result at fixed T against coupling and at fixed coupling against T respectively. We again see good agreement at weak coupling and high temperature as expected, although the deviation occurs at lower $\lambda$. This can easily be understood from the definition of $X = s_{1}+s_{2}=\sqrt{2}d_{+}$, so the reservoir couples to the bright mode with strength $\sqrt{2}\lambda$, meaning deviations set in at smaller $\lambda$ than for the RLM. The coherence also remains finite as $T\to 0$, consistent with the persistence of coherences at low temperatures found in Sec.~\ref{sec:SM-lowT}. Figures~\ref{fig:two-mode}(b), (d) show the associated comparison but now for the full state in terms of the trace distance $\mathcal{T}(\rho,\sigma) = \frac{1}{2}\Trs\sqrt{(\rho-\sigma)^{\dagger}(\rho-\sigma)}$, with $\tau_{\textrm{MF}}^{(2)}$ against the exact result, in addition to the difference between the bare Gibbs and the exact result. The behaviour is qualitatively the same for both the trace distance and the coherence.

\textit{Single impurity Anderson model} - As a final example, we consider the paradigmatic single impurity Anderson model (SIAM)~\cite{Anderson1961}, with Hamiltonian
\begin{equation}
H_S=\varepsilon_d\,(n_\uparrow+n_\downarrow)+U\,n_\uparrow n_\downarrow,
\qquad n_\sigma=d_\sigma^{\dagger}d_\sigma,\label{eq:examples_8}
\end{equation}
coupled to a spinful reservoir,
\begin{align}
H&=H_S+\sum_{\sigma}\int\! d\varepsilon\,\varepsilon\,a^{\dagger}_{\varepsilon\sigma}a_{\varepsilon\sigma} \nonumber \\
&+\lambda\sum_{\sigma}\int\! d\varepsilon\,\sqrt{J(\varepsilon)}\big(d_\sigma^{\dagger}a_{\varepsilon\sigma}+a^{\dagger}_{\varepsilon\sigma}d_\sigma\big).
\label{eq:examples_9}
\end{align}
We take $\sigma\in\{\uparrow,\downarrow\}$ to label spin, where $\bar\sigma$ denotes the opposite spin. Unlike the previous two examples, this model is interacting, so the exact state is not accessible via the correlation-matrix approach. It also involves two reservoirs, one for each spin, coupled through $X_{\sigma}=d_{\sigma}$ and $E_{\sigma}=\int d\varepsilon \sqrt{J(\varepsilon)}a_{\varepsilon\sigma}$. However, as the two spin reservoirs are independent, Eq.~\eqref{eq:result_2} holds with the correction summed over $\sigma$ (see Sec.~\ref{sec:SM-SIAM}). Each spin operator decomposes into two eigenoperators $X_{\sigma,\omega_1}=d_\sigma(1-n_{\bar\sigma}),
X_{\sigma,\omega_2}=d_\sigma n_{\bar\sigma}$
so that $X_\sigma=X_{\sigma,\omega_1}+X_{\sigma,\omega_2}$ where $\omega_1\equiv\mu-\varepsilon_d,\omega_2\equiv\mu-\varepsilon_d-U$. These two Bohr frequencies correspond to adding a particle to the empty and to the singly occupied level respectively. Defining the Fock states $\ket{0}, \ket{\sigma}=d_\sigma^{\dagger}\ket{0},\ket{2}=d_\uparrow^{\dagger}d_\downarrow^{\dagger}\ket{0}$, Eq.~\eqref{eq:result_2} gives
\begin{equation}
\tau_{\MF}^{(2)}=\sum_{k}p_k^{(0)}\Big[1+\lambda^2\big(w_k-\bar w\big)\Big]\ket{k}\bra{k},
\quad
\bar w\equiv\sum_kp_k^{(0)}w_k,\label{eq:examples_10}
\end{equation}
\begin{figure}[t!]
  \centering
  \includegraphics[width=0.49\textwidth]{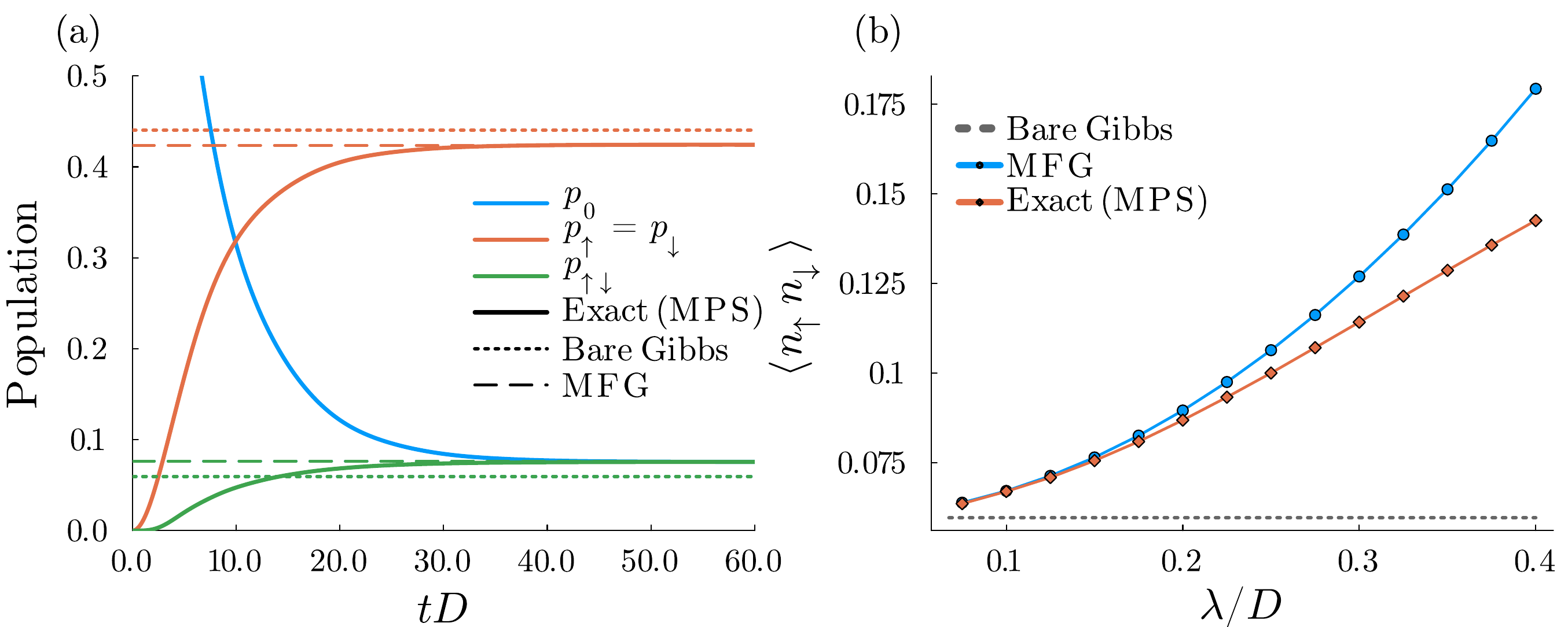}
  \caption{Comparison between the MFG prediction Eq.~\eqref{eq:result_2}, the bare Gibbs state and the numerically exact value from MPS real time evolution. (a) Time evolution of the four diagonal components of the impurity state with initial state $\rho_0=\ket{0}\bra{0}$, at coupling $\lambda=0.15D$. (b) Steady state double occupancy of the MFG prediction compared to the converged values from the MPS calculation and the bare Gibbs state. Parameters: $\beta=5/D,\varepsilon_{d}=-0.4D,U=0.8D, \mu=0$.}
  \label{fig:SIAM}
\end{figure}
for $k\in\{0,\uparrow,\downarrow,2\}$, where $p_k^{(0)}$ are the bare Gibbs state occupations. For the definitions of $w_k$ in terms of $D_1,D_2$ see Sec.~\ref{sec:SM-SIAM}. As required by charge and $S_z$ conservation, the state $\tau_{\MF}^{(2)}$ is diagonal. Fig.~\ref{fig:SIAM}(b) shows the double occupation as predicted by Eq.~\eqref{eq:examples_10} and the converged steady state of the MPS calculation against $\lambda$ for the particle hole symmetric case $\varepsilon_{d}=-U/2$ at strong interaction $U$. We see good agreement in the expected regime of weak to intermediate coupling. The Hamiltonian of mean force can be written analogously to the bare Hamiltonian $H_{S}$ with effective energies and interaction $\tilde{\varepsilon}_{d}$, $\tilde{U}$ (see Sec.~\ref{sec:SM-SIAM}). Fig.~\ref{fig:SIAM}(a) shows the time evolution of the four diagonal components of the impurity. Together, these results demonstrate the accuracy and validity of our result in the expected parameter regime.

\section{Conclusion}

We have derived the grand-canonical MFG state to second order in the coupling for an arbitrary fermionic system exchanging particles with a fermionic reservoir, together with the corresponding Hamiltonian of mean force, for which the MFG state takes a Gibbs-like form. The correction separates into particle and hole channels weighted by the empty and filled reservoir states, and the KMS condition fixes all coefficients in terms of a single reservoir function. As in the bosonic case~\cite{PhysRevLett.127.250601}, the MFG state generically carries coherences in the bare energy eigenbasis, where we found for two independent modes these are mediated entirely through the reservoir and persist at $T=0$, giving the fermionic analogue to~\cite{LewisWilliamson2026}. We compared with exact results for the quadratic models and with thermofield MPS dynamics for the single impurity Anderson model, finding good agreement at weak to intermediate coupling as expected. Our results add to the literature on the MFG state which has thus far been mostly restricted to the bosonic regime. This provides a reference state needed to define strong-coupling thermodynamic quantities for fermionic systems such as quantum dots~\cite{Josefsson2018,Burkard2023}, molecular junctions~\cite{CuevasScheer2017,ThossEvers2018} and thermal machines~\cite{Brenes2020}. A natural extension would be to explore the possibility of tractable expressions for the fermionic MFG beyond the perturbative coupling regime, although this is a challenging problem, especially as there exists no fermionic analogue of the polaron transform which is the usual route taken for bosonic systems~\cite{10.1116/5.0073853}. One possible approach would be to use the reaction-coordinate mapping~\cite{Strasberg2018FermionicRC} to treat the strong system-reservoir coupling exactly and use Eq.~\eqref{eq:result_2} on the residual reservoir coupling, although we leave this for future work.

\begin{acknowledgments}
D.J.S acknowledges support from EPSRC Grant EP/Z534936/1.

\end{acknowledgments}

\bibliography{References.bib}

@article{PhysRevLett.127.250601,
  title = {Weak and Ultrastrong Coupling Limits of the Quantum Mean Force Gibbs State},
  author = {Cresser, J. D. and Anders, J.},
  journal = {Phys. Rev. Lett.},
  volume = {127},
  issue = {25},
  pages = {250601},
  numpages = {7},
  year = {2021},
  month = {Dec},
  publisher = {American Physical Society},
  doi = {10.1103/PhysRevLett.127.250601},
  url = {https://link.aps.org/doi/10.1103/PhysRevLett.127.250601}
}

@article{10.1116/5.0073853,
  author    = {Trushechkin, A. S. and Merkli, M. and Cresser, J. D. and Anders, J.},
  title     = {Open quantum system dynamics and the mean force {G}ibbs state},
  journal   = {AVS Quantum Science},
  volume    = {4},
  number    = {1},
  pages     = {012301},
  year      = {2022},
  month     = mar,
  doi       = {10.1116/5.0073853},
  url       = {https://doi.org/10.1116/5.0073853},
  eprint    = {2110.01671},
  archivePrefix = {arXiv}
}

@article{RevModPhys.92.041002,
  title = {Colloquium: Statistical mechanics and thermodynamics at strong coupling: Quantum and classical},
  author = {Talkner, Peter and H\"anggi, Peter},
  journal = {Rev. Mod. Phys.},
  volume = {92},
  issue = {4},
  pages = {041002},
  numpages = {26},
  year = {2020},
  month = {Oct},
  publisher = {American Physical Society},
  doi = {10.1103/RevModPhys.92.041002},
  url = {https://link.aps.org/doi/10.1103/RevModPhys.92.041002}
}

@article{PhysRevX.7.011008,
  title = {Stochastic and Macroscopic Thermodynamics of Strongly Coupled Systems},
  author = {Jarzynski, Christopher},
  journal = {Phys. Rev. X},
  volume = {7},
  issue = {1},
  pages = {011008},
  numpages = {20},
  year = {2017},
  month = {Jan},
  publisher = {American Physical Society},
  doi = {10.1103/PhysRevX.7.011008},
  url = {https://link.aps.org/doi/10.1103/PhysRevX.7.011008}
}

@article{Purkayastha2020,
  author    = {Purkayastha, Archak and Guarnieri, Giacomo and Mitchison, Mark T. and Filip, Radim and Goold, John},
  title     = {Tunable phonon-induced steady-state coherence in a double-quantum-dot charge qubit},
  journal   = {npj Quantum Information},
  volume    = {6},
  number    = {1},
  pages     = {27},
  year      = {2020},
  month     = mar,
  publisher = {Nature Publishing Group},
  doi       = {10.1038/s41534-020-0256-6},
  url       = {https://doi.org/10.1038/s41534-020-0256-6}
}

@article{Seifert2016,
  title = {First and Second Law of Thermodynamics at Strong Coupling},
  author = {Seifert, Udo},
  journal = {Phys. Rev. Lett.},
  volume = {116},
  issue = {2},
  pages = {020601},
  numpages = {5},
  year = {2016},
  month = {Jan},
  publisher = {American Physical Society},
  doi = {10.1103/PhysRevLett.116.020601},
  url = {https://link.aps.org/doi/10.1103/PhysRevLett.116.020601}
}

@article{Rivas2020,
  title = {Strong Coupling Thermodynamics of Open Quantum Systems},
  author = {Rivas, \'Angel},
  journal = {Phys. Rev. Lett.},
  volume = {124},
  issue = {16},
  pages = {160601},
  numpages = {7},
  year = {2020},
  month = {Apr},
  publisher = {American Physical Society},
  doi = {10.1103/PhysRevLett.124.160601},
  url = {https://link.aps.org/doi/10.1103/PhysRevLett.124.160601}
}

@incollection{Miller2018,
  author    = {Harry J. D. Miller},
  title     = {Hamiltonian of Mean Force for Strongly-Coupled Systems},
  booktitle = {Thermodynamics in the Quantum Regime: Fundamental Aspects and New Directions},
  editor    = {Felix Binder and Luis A. Correa and Christian Gogolin and Janet Anders and Gerardo Adesso},
  series    = {Fundamental Theories of Physics},
  volume    = {195},
  pages     = {531--549},
  publisher = {Springer},
  address   = {Cham},
  year      = {2019},
  doi       = {10.1007/978-3-319-99046-0_22}
}

@article{PhysRevE.95.062101,
  title = {Stochastic thermodynamics in the strong coupling regime: An unambiguous approach based on coarse graining},
  author = {Strasberg, Philipp and Esposito, Massimiliano},
  journal = {Phys. Rev. E},
  volume = {95},
  issue = {6},
  pages = {062101},
  numpages = {9},
  year = {2017},
  month = {Jun},
  publisher = {American Physical Society},
  doi = {10.1103/PhysRevE.95.062101},
  url = {https://link.aps.org/doi/10.1103/PhysRevE.95.062101}
}

@article{Thingna2012,
  author  = {J. Thingna and J.-S. Wang and P. H\"anggi},
  title   = {Generalized Gibbs state with modified Redfield solution: Exact agreement up to second order},
  journal = {Journal of Chemical Physics},
  volume  = {136},
  pages   = {194110},
  year    = {2012},
  doi     = {10.1063/1.4714558}
}

@article{Geva2000,
  author  = {Geva, Eitan and Rosenman, Efrat and Tannor, David},
  title   = {On the second-order corrections to the quantum canonical equilibrium density matrix},
  journal = {The Journal of Chemical Physics},
  volume  = {113},
  number  = {4},
  pages   = {1380--1390},
  year    = {2000},
  doi     = {10.1063/1.481928}
}

@article{MoriMiyashita2008,
  author  = {Takashi Mori and Seiji Miyashita},
  title   = {Dynamics of the Density Matrix in Contact with a Thermal Bath and the Quantum Master Equation},
  journal = {Journal of the Physical Society of Japan},
  volume  = {77},
  pages   = {124005},
  year    = {2008},
  doi     = {10.1143/JPSJ.77.124005}
}

@article{Chiu2022,
  author  = {Yiu-Fung Chiu and Aidan Strathearn and Jonathan Keeling},
  title   = {Numerical evaluation and robustness of the quantum mean force Gibbs state},
  journal = {Phys. Rev. A},
  volume  = {106},
  issue    = {1},
  pages   = {012204},
  year    = {2022},
  doi     = {10.1103/PhysRevA.106.012204}
}

@article{Cerisola2024,
  author  = {Francesco Cerisola and Marco Berritta and Simone Scali and
             Samuel A. R. Horsley and James D. Cresser and Janet Anders},
  title   = {Quantum--classical correspondence in spin--boson equilibrium states at arbitrary coupling},
  journal = {New Journal of Physics},
  volume  = {26},
  pages   = {053032},
  year    = {2024},
  doi     = {10.1088/1367-2630/ad4818}
}

@article{AntoSztrikacs2023,
  title = {Effective-Hamiltonian Theory of Open Quantum Systems at Strong Coupling},
  author = {Anto-Sztrikacs, Nicholas and Nazir, Ahsan and Segal, Dvira},
  journal = {PRX Quantum},
  volume = {4},
  issue = {2},
  pages = {020307},
  numpages = {35},
  year = {2023},
  month = {Apr},
  publisher = {American Physical Society},
  doi = {10.1103/PRXQuantum.4.020307},
  url = {https://link.aps.org/doi/10.1103/PRXQuantum.4.020307}
}

@article{KumarGhosh2024,
  author  = {Abhishek Kumar and Ayan Ghosh},
  title   = {Ultrastrong coupling limit to quantum mean force Gibbs state for anharmonic environment},
  journal = {Journal of Chemical Physics},
  volume  = {161},
  pages   = {174111},
  year    = {2024},
  doi     = {10.1063/5.0223734}
}

@article{Guarnieri2018,
  author  = {Giovanni Guarnieri and Michal Kol{\'a}{\v r} and Radim Filip},
  title   = {Steady-State Coherences by Composite System-Bath Interactions},
  journal = {Phys. Rev. Lett.},
  volume  = {121},
  pages   = {070401},
  year    = {2018},
  doi     = {10.1103/PhysRevLett.121.070401}
}

@misc{LewisWilliamson2026,
  author       = {Lewis A. Williamson and William McEniery and Francesco Cerisola and Janet Anders},
  title        = {Reservoir-mediated spin entanglement in the mean-force Gibbs state},
  year         = {2026},
  eprint       = {2604.26562},
  archivePrefix= {arXiv},
  primaryClass = {quant-ph}
}

@article{Haegeman2016,
  author  = {Jutho Haegeman and Christian Lubich and Ivan Oseledets and Bart Vandereycken and Frank Verstraete},
  title   = {Unifying time evolution and optimization with matrix product states},
  journal = {Physical Review B},
  volume  = {94},
  pages   = {165116},
  year    = {2016},
  doi     = {10.1103/PhysRevB.94.165116}
}

@article{Anderson1961,
  author  = {P. W. Anderson},
  title   = {Localized Magnetic States in Metals},
  journal = {Physical Review},
  volume  = {124},
  pages   = {41--53},
  year    = {1961},
  doi     = {10.1103/PhysRev.124.41}
}

@article{Ochoa2016,
  title = {Energy distribution and local fluctuations in strongly coupled open quantum systems: The extended resonant level model},
  author = {Ochoa, Maicol A. and Bruch, Anton and Nitzan, Abraham},
  journal = {Phys. Rev. B},
  volume = {94},
  issue = {3},
  pages = {035420},
  numpages = {7},
  year = {2016},
  month = {Jul},
  publisher = {American Physical Society},
  doi = {10.1103/PhysRevB.94.035420},
  url = {https://link.aps.org/doi/10.1103/PhysRevB.94.035420}
}

@article{Dou2018,
  title = {Universal approach to quantum thermodynamics in the strong coupling regime},
  author = {Dou, Wenjie and Ochoa, Maicol A. and Nitzan, Abraham and Subotnik, Joseph E.},
  journal = {Phys. Rev. B},
  volume = {98},
  issue = {13},
  pages = {134306},
  numpages = {15},
  year = {2018},
  month = {Oct},
  publisher = {American Physical Society},
  doi = {10.1103/PhysRevB.98.134306},
  url = {https://link.aps.org/doi/10.1103/PhysRevB.98.134306}
}

@article{Horodecki2013,
  author  = {Micha{\l} Horodecki and Jonathan Oppenheim},
  title   = {Fundamental limitations for quantum and nanoscale thermodynamics},
  journal = {Nature Communications},
  volume  = {4},
  pages   = {2059},
  year    = {2013},
  doi     = {10.1038/ncomms3059}
}

@article{Brandao2015,
  author  = {Fernando G. S. L. Brand{\~a}o and Michal Horodecki and Jonathan Oppenheim and Joseph M. Renes and Robert W. Spekkens},
  title   = {The second laws of quantum thermodynamics},
  journal = {Proceedings of the National Academy of Sciences},
  volume  = {112},
  number  = {11},
  pages   = {3275--3279},
  year    = {2015},
  doi     = {10.1073/pnas.1411728112}
}

@article{Campaioli2018,
  title = {Enhancing the Charging Power of Quantum Batteries},
  author = {Campaioli, Francesco and Pollock, Felix A. and Binder, Felix C. and C{\'e}leri, Lucas and Goold, John and Vinjanampathy, Sai and Modi, Kavan},
  journal = {Phys. Rev. Lett.},
  volume = {118},
  issue = {15},
  pages = {150601},
  numpages = {6},
  year = {2017},
  month = {Apr},
  publisher = {American Physical Society},
  doi = {10.1103/PhysRevLett.118.150601},
  url = {https://link.aps.org/doi/10.1103/PhysRevLett.118.150601}
}

@article{Bhattacharjee2021,
  author  = {Sourav Bhattacharjee and Amit Dutta},
  title   = {Quantum thermal machines and batteries},
  journal = {European Physical Journal B},
  volume  = {94},
  pages   = {239},
  year    = {2021},
  doi     = {10.1140/epjb/s10051-021-00345-7}
}

@article{RevModPhys.83.771,
  title = {Colloquium: Quantum fluctuation relations: Foundations and applications},
  author = {Campisi, Michele and H\"anggi, Peter and Talkner, Peter},
  journal = {Rev. Mod. Phys.},
  volume = {83},
  issue = {3},
  pages = {771--791},
  numpages = {0},
  year = {2011},
  month = {Jul},
  publisher = {American Physical Society},
  doi = {10.1103/RevModPhys.83.771},
  url = {https://link.aps.org/doi/10.1103/RevModPhys.83.771}
}

@article{Kosloff2019,
  title   = {Quantum thermodynamics and open-systems modeling},
  author  = {Kosloff, Ronnie},
  journal = {The Journal of Chemical Physics},
  volume  = {150},
  number  = {20},
  pages   = {204105},
  year    = {2019},
  doi     = {10.1063/1.5096173},
}

@article{PhysRevA.43.2046,
  title = {Quantum statistical mechanics in a closed system},
  author = {Deutsch, J. M.},
  journal = {Phys. Rev. A},
  volume = {43},
  issue = {4},
  pages = {2046--2049},
  numpages = {0},
  year = {1991},
  month = {Feb},
  publisher = {American Physical Society},
  doi = {10.1103/PhysRevA.43.2046},
  url = {https://link.aps.org/doi/10.1103/PhysRevA.43.2046}
}

@article{PhysRevE.50.888,
  title = {Chaos and quantum thermalization},
  author = {Srednicki, Mark},
  journal = {Phys. Rev. E},
  volume = {50},
  issue = {2},
  pages = {888--901},
  numpages = {0},
  year = {1994},
  month = {Aug},
  publisher = {American Physical Society},
  doi = {10.1103/PhysRevE.50.888},
  url = {https://link.aps.org/doi/10.1103/PhysRevE.50.888}
}

@article{Amosov_2016,
   title={{Spectral} properties of reduced fermionic density operators and parity superselection rule},
   volume={16},
   ISSN={1573-1332},
   url={https://link.springer.com/article/10.1007/s11128-016-1467-9},
   DOI={10.1007/s11128-016-1467-9},
   number={1},
   journal={Quantum Information Processing},
   publisher={Springer Science and Business Media LLC},
   author={Amosov, Grigori G. and Filippov, Sergey N.},
   year={2016},
   month=dec }

@article{PhysRevD.1.3267,
  title = {{Superselection} {Rule} for {Charge}},
  author = {Wick, G. C. and Wightman, A. S. and Wigner, Eugene P.},
  journal = {Phys. Rev. D},
  volume = {1},
  issue = {12},
  pages = {3267--3269},
  numpages = {0},
  year = {1970},
  month = {Jun},
  publisher = {American Physical Society},
  doi = {10.1103/PhysRevD.1.3267},
  url = {https://link.aps.org/doi/10.1103/PhysRevD.1.3267}
}

@article{PhysRev.88.101,
  title = {{The} {Intrinsic} {Parity} of {Elementary} {Particles}},
  author = {Wick, G. C. and Wightman, A. S. and Wigner, E. P.},
  journal = {Phys. Rev.},
  volume = {88},
  issue = {1},
  pages = {101--105},
  numpages = {0},
  year = {1952},
  month = {Oct},
  publisher = {American Physical Society},
  doi = {10.1103/PhysRev.88.101},
  url = {https://link.aps.org/doi/10.1103/PhysRev.88.101}
}

@article{PhysRevB.69.075111,
  title = {Many-body density matrices for free fermions},
  author = {Cheong, Siew-Ann and Henley, Christopher L.},
  journal = {Phys. Rev. B},
  volume = {69},
  issue = {7},
  pages = {075111},
  numpages = {12},
  year = {2004},
  month = {Feb},
  publisher = {American Physical Society},
  doi = {10.1103/PhysRevB.69.075111},
  url = {https://link.aps.org/doi/10.1103/PhysRevB.69.075111}
}

@article{TimofeevTrushechkin2022,
  author    = {Timofeev, G. M. and Trushechkin, A. S.},
  title     = {Hamiltonian of mean force in the weak-coupling and high-temperature approximations and refined quantum master equations},
  journal   = {International Journal of Modern Physics A},
  volume    = {37},
  number    = {20n21},
  pages     = {2243021},
  year      = {2022},
  publisher = {World Scientific Publishing Company},
  doi       = {10.1142/S0217751X22430217},
  url       = {https://doi.org/10.1142/S0217751X22430217}
}

@article{Josefsson2018,
  author  = {Josefsson, M. and Svilans, A. and Burke, A. M. and Hoffmann, E. A. and Fahlvik, S. and Thelander, C. and Leijnse, M. and Linke, H.},
  title   = {A quantum-dot heat engine operating close to the thermodynamic efficiency limits},
  journal = {Nat. Nanotechnol.},
  volume  = {13},
  pages   = {920},
  year    = {2018},
  doi     = {10.1038/s41565-018-0200-5}
}

@article{KumarAthulyaGhosh2025,
  author  = {Kumar, P. and Athulya, K. P. and Ghosh, S.},
  title   = {Equivalence between the second order steady state for the spin-boson model and its quantum mean force {G}ibbs state},
  journal = {Phys. Rev. B},
  volume  = {111},
  pages   = {115423},
  year    = {2025},
  doi     = {10.1103/PhysRevB.111.115423}
}

@article{Nusseler2020,
  author  = {N{\"u}{\ss}eler, A. and Dhand, I. and Huelga, S. F. and Plenio, M. B.},
  title   = {Efficient simulation of open quantum systems coupled to a fermionic bath},
  journal = {Phys. Rev. B},
  volume  = {101},
  pages   = {155134},
  year    = {2020},
  doi     = {10.1103/PhysRevB.101.155134}
}

@article{Peschel2003,
  author  = {Peschel, I.},
  title   = {Calculation of reduced density matrices from correlation functions},
  journal = {J. Phys. A: Math. Gen.},
  volume  = {36},
  pages   = {L205},
  year    = {2003},
  doi     = {10.1088/0305-4470/36/14/101}
}

@article{deVegaBanuls2015,
  author  = {de Vega, I. and Ba{\~n}uls, M.-C.},
  title   = {Thermofield-based chain-mapping approach for open quantum systems},
  journal = {Phys. Rev. A},
  volume  = {92},
  pages   = {052116},
  year    = {2015},
  doi     = {10.1103/PhysRevA.92.052116}
}

@article{RevModPhys.89.041003,
  title = {Colloquium: Quantum coherence as a resource},
  author = {Streltsov, Alexander and Adesso, Gerardo and Plenio, Martin B.},
  journal = {Rev. Mod. Phys.},
  volume = {89},
  issue = {4},
  pages = {041003},
  numpages = {34},
  year = {2017},
  month = {Oct},
  publisher = {American Physical Society},
  doi = {10.1103/RevModPhys.89.041003},
  url = {https://link.aps.org/doi/10.1103/RevModPhys.89.041003}
}

@article{PhysRevLett.122.110601,
  title = {Experimental Demonstration of Quantum Effects in the Operation of Microscopic Heat Engines},
  author = {Klatzow, James and Becker, Jonas N. and Ledingham, Patrick M. and Weinzetl, Christian and Kaczmarek, Krzysztof T. and Saunders, Dylan J. and Nunn, Joshua and Walmsley, Ian A. and Uzdin, Raam and Poem, Eilon},
  journal = {Phys. Rev. Lett.},
  volume = {122},
  issue = {11},
  pages = {110601},
  numpages = {6},
  year = {2019},
  month = {Mar},
  publisher = {American Physical Society},
  doi = {10.1103/PhysRevLett.122.110601},
  url = {https://link.aps.org/doi/10.1103/PhysRevLett.122.110601}
}

@article{PhysRevA.87.022338,
  title = {Fermionic-mode entanglement in quantum information},
  author = {Friis, Nicolai and Lee, Antony R. and Bruschi, David Edward},
  journal = {Phys. Rev. A},
  volume = {87},
  issue = {2},
  pages = {022338},
  numpages = {7},
  year = {2013},
  month = {Feb},
  publisher = {American Physical Society},
  doi = {10.1103/PhysRevA.87.022338},
  url = {https://link.aps.org/doi/10.1103/PhysRevA.87.022338}
}

@article{Suzuki1977,
  author  = {Suzuki, Masuo},
  title   = {On the convergence of exponential operators---the {Zassenhaus} formula, {BCH} formula and systematic approximants},
  journal = {Communications in Mathematical Physics},
  volume  = {57},
  pages   = {193--200},
  year    = {1977}
}

@article{PhysRevLett.81.2594,
  title = {Decoherence-Free Subspaces for Quantum Computation},
  author = {Lidar, D. A. and Chuang, I. L. and Whaley, K. B.},
  journal = {Phys. Rev. Lett.},
  volume = {81},
  issue = {12},
  pages = {2594--2597},
  numpages = {0},
  year = {1998},
  month = {Sep},
  publisher = {American Physical Society},
  doi = {10.1103/PhysRevLett.81.2594},
  url = {https://link.aps.org/doi/10.1103/PhysRevLett.81.2594}
}

@article{Buca2019,
  author  = {Bu{\v{c}}a, Berislav and Tindall, Joseph and Jaksch, Dieter},
  title   = {Non-stationary coherent quantum many-body dynamics through dissipation},
  journal = {Nature Communications},
  volume  = {10},
  number  = {1},
  pages   = {1730},
  year    = {2019},
  doi     = {10.1038/s41467-019-09757-y}
}

@book{CuevasScheer2017,
  author    = {Cuevas, J. C. and Scheer, E.},
  title     = {Molecular Electronics: An Introduction to Theory and Experiment},
  edition   = {2nd}, publisher = {World Scientific}, address = {Singapore}, year = {2017}
}

@article{ThossEvers2018,
  author  = {Thoss, M. and Evers, F.},
  title   = {Perspective: Theory of quantum transport in molecular junctions},
  journal = {J. Chem. Phys.}, volume = {148}, pages = {030901}, year = {2018},
  doi     = {10.1063/1.5003306}
}

@article{Burkard2023,
  author  = {Burkard, Guido and Ladd, Thaddeus D. and Pan, Andrew and Nichol, John M. and Petta, Jason R.},
  title   = {Semiconductor spin qubits},
  journal = {Rev. Mod. Phys.}, volume = {95}, pages = {025003}, year = {2023},
  doi     = {10.1103/RevModPhys.95.025003}
}

@article{Brenes2020,
  author  = {Brenes, Marlon and Mendoza-Arenas, Juan Jos{\'e} and Purkayastha, Archak and Mitchison, Mark T. and Clark, Stephen R. and Goold, John},
  title   = {Tensor-network method to simulate strongly interacting quantum thermal machines},
  journal = {Phys. Rev. X}, volume = {10}, pages = {031040}, year = {2020},
  doi     = {10.1103/PhysRevX.10.031040}
}

@article{Strasberg2018FermionicRC,
  title         = {Fermionic reaction coordinates and their application to an autonomous {Maxwell} demon in the strong-coupling regime},
  author        = {Strasberg, Philipp and Schaller, Gernot and Schmidt, Thomas L. and Esposito, Massimiliano},
  journal       = {Phys. Rev. B},
  volume        = {97},
  issue         = {20},
  pages         = {205405},
  year          = {2018},
  doi           = {10.1103/PhysRevB.97.205405},
  eprint        = {1711.08914},
  archivePrefix = {arXiv}
}

@article{Strachan2024,
  author  = {Strachan, David J. and Purkayastha, Archak and Clark, Stephen R.},
  title   = {Extracting dynamical maps of non-{M}arkovian open quantum systems},
  journal = {The Journal of Chemical Physics},
  volume  = {161},
  number  = {15},
  pages   = {154105},
  year    = {2024},
  doi     = {10.1063/5.0228428},
  eprint  = {2409.17051},
  archivePrefix = {arXiv}
}

@article{PhysRevB.104.014303,
  title = {Efficient mapping for Anderson impurity problems with matrix product states},
  author = {Kohn, Lucas and Santoro, Giuseppe E.},
  journal = {Phys. Rev. B},
  volume = {104},
  issue = {1},
  pages = {014303},
  numpages = {6},
  year = {2021},
  month = {Jul},
  publisher = {American Physical Society},
  doi = {10.1103/PhysRevB.104.014303},
  url = {https://link.aps.org/doi/10.1103/PhysRevB.104.014303}
}

@article{Chin2010,
    author = {Chin, Alex W. and Rivas, \'Angel and Huelga, Susana F. and Plenio, Martin B.},
    title = "{{Exact} mapping between system-reservoir quantum models and semi-infinite discrete chains using orthogonal polynomials}",
    journal = {J. Math. Phys.},
    volume = {51},
    number = {9},
    pages = {092109},
    year = {2010},
    month = {09},
    issn = {0022-2488},
    doi = {10.1063/1.3490188},
    url = {https://doi.org/10.1063/1.3490188},
}

@article{Fishman_2022,
	doi = {10.21468/scipostphyscodeb.4},
  
	url = {https://doi.org/10.21468%2Fscipostphyscodeb.4},
  
	year = 2022,
	month = {aug},
  
	publisher = {Stichting {SciPost}
},
  
	author = {Matthew Fishman and Steven White and Edwin Stoudenmire},
  
	title = {{The} {{ITensor}} Software Library for Tensor Network Calculations},
  
	journal = {{SciPost} Physics Codebases}
}

@article{PhysRevB.92.155126,
  title = {How to discretize a quantum bath for real-time evolution},
  author = {de Vega, In{\'e}s and Schollw{\"o}ck, Ulrich and Wolf, F. Alexander},
  journal = {Phys. Rev. B},
  volume = {92},
  issue = {15},
  pages = {155126},
  numpages = {14},
  year = {2015},
  month = {Oct},
  publisher = {American Physical Society},
  doi = {10.1103/PhysRevB.92.155126},
  url = {https://link.aps.org/doi/10.1103/PhysRevB.92.155126}
}

@article{RevModPhys.96.031001,
  title = {Colloquium: Quantum batteries},
  author = {Campaioli, Francesco and Gherardini, Stefano and Quach, James Q. and Polini, Marco and Andolina, Gian Marcello},
  journal = {Rev. Mod. Phys.},
  volume = {96},
  issue = {3},
  pages = {031001},
  numpages = {30},
  year = {2024},
  month = {Jul},
  publisher = {American Physical Society},
  doi = {10.1103/RevModPhys.96.031001},
  url = {https://link.aps.org/doi/10.1103/RevModPhys.96.031001}
}

@article{Uusnakki2026,
  author  = {Uusn{\"a}kki, Tuomas and M{\"o}rstedt, Timm and Teixeira, Wallace and Rasola, Miika and M{\"o}tt{\"o}nen, Mikko},
  title   = {Initial demonstration of a quantum heat engine based on dissipation-engineered superconducting circuits},
  journal = {Nature Communications},
  volume  = {17},
  pages   = {6054},
  year    = {2026},
  doi     = {10.1038/s41467-026-72651-x},
  url     = {https://www.nature.com/articles/s41467-026-72651-x}
}
\clearpage


\onecolumngrid
\begin{center}
  \textbf{\large Supplementary Material for: Second-order mean force Gibbs state of fermionic open systems}\\[.2cm]
  David J. Strachan$^{1}$\\[.1cm]
  {\itshape ${}^1$H. H. Wills Physics Laboratory, University of Bristol, Bristol BS8 1TL, United Kingdom}\\
(Dated: \today)\\[1cm]
\end{center}

\setcounter{equation}{0}
\setcounter{figure}{0}
\setcounter{table}{0}
\setcounter{section}{0}

\renewcommand{\theequation}{S\arabic{equation}}
\renewcommand{\thefigure}{S\arabic{figure}}
\renewcommand{\thetable}{S\Roman{table}}
\renewcommand{\thesection}{S\Roman{section}}
\renewcommand{\bibnumfmt}[1]{[S#1]}

\section{Derivation of the main result}
\label{sec:SM-derivation}

Here we give a detailed derivation of Eq.~\eqref{eq:result_2} at arbitrary chemical potential $\mu$. We consider a system and environment with total Hamiltonian
\begin{equation}
H = H_{S}+H_{R}+\lambda V,
\label{eq:SM_derivation_1}
\end{equation}
where $H_{S}$ and $H_{R}$ are the system and reservoir Hamiltonians and $\lambda V$ their interaction, with $\lambda$ the coupling parameter. In the bosonic literature $V$ is usually taken to be of the factorised form $V=X \otimes B$ with $X$ and $B$ Hermitian system and reservoir operators~\cite{PhysRevLett.127.250601,10.1116/5.0073853}.
Here we instead take
\begin{equation}
V = X^{\dagger}E+E^{\dagger}X,
\label{eq:SM_derivation_2}
\end{equation}
which describes an exchange, or hopping, interaction, with $X$ and $E$ being odd-parity system and reservoir operators respectively. Importantly, this does not assume a tensor product structure as fermions do not possess one. This is the generic coupling for fermionic reservoirs, and is also a legitimate choice for bosonic ones i.e. the Jaynes--Cummings interaction. Specifically we consider the canonical setup as outlined in section~\ref{sec:Setup} in which we have
\begin{equation} \label{eq:SM_derivation_3}
E = \int\! d\varepsilon\,\sqrt{J(\varepsilon)}\,a_{\varepsilon}.
\end{equation}
We assume that the coupling conserves the total particle number $N=N_S+N_R$, with $N_R=\int\! d\varepsilon\, a^{\dagger}_{\varepsilon}a_{\varepsilon}$, and that $[H_S,N_S]=0$. Since $[N_R,E]=-E$, number conservation of Eq.~\eqref{eq:SM_derivation_2} requires
\begin{equation}
[N_S,X]=-X ,
\label{eq:SM_derivation_4}
\end{equation}
i.e.\ $X$ removes exactly one particle from the system, so that $[V,N]=0$ and $[H,N]=0$. The global state is then the grand canonical Gibbs state $\tau_{\rm tot} = e^{-\beta(H-\mu N)}/\Trs(e^{-\beta(H-\mu N)})$, and its generator can be written as
\begin{equation}
H-\mu N = K_S+K_R+\lambda V\equiv K_0+\lambda V,
\qquad
K_S\equiv H_S-\mu N_S,
\qquad
K_R\equiv\int\! d\varepsilon\,\xi\, a^{\dagger}_{\varepsilon}a_{\varepsilon},
\qquad
\xi\equiv\varepsilon-\mu .
\label{eq:SM_derivation_5}
\end{equation}
The system Gibbs state is $\tau_S=e^{-\beta K_S}/Z_S$ with $Z_S=\Trs_S(e^{-\beta K_S})$, and the reservoir Gibbs state is $\tau_R=e^{-\beta K_R}/Z_R$.
\subsection{Fermionic partial trace and parity} \label{app:fermionic partial trace}

For fermions, the partial trace is somewhat ambiguous due to the lack of a tensor product structure of the fermionic Fock space. Fermionic Fock spaces are defined by the algebra of canonical anticommutation relations (CAR) $\mathcal{A}$, defined by
\begin{equation}  \label{eq:FPT1}
\{b_{i}^{\dag},b_{j}\} = \delta_{ij},
\end{equation}
\begin{equation}  \label{eq:FPT2}
\{b_{i},b_{j}\} = \{b_{i}^{\dag},b_{j}^{\dag}\} = 0,
\end{equation}
where $b_i^{\dag}$ ($b_i$) are creation (annihilation) operators. Fock states are then defined through a specific ordering of mode operators,
\begin{equation} \label{eq:FPT3}
\ket{j_{1},...,j_{n}} \equiv (b_{1}^{\dag})^{j_1}(b_{2}^{\dag})^{j_2}...(b_{n}^{\dag})^{j_n}\ket{\textrm{vac}}.
\end{equation}  
For bosons this ordering is arbitrary as creation operators of different modes commute, but for fermions this is not the case. It is clear from this that $\ket{j_1,...,j_n}$ cannot be written as a tensor product $\ket{j_1}\otimes...\otimes\ket{j_n}$ due to this anticommutativity, and as such the standard partial trace definition in terms of such a tensor product does not directly apply. We can instead define a fermionic partial trace, or mode reduced state~\cite{PhysRevA.87.022338,Amosov_2016} $\rho_{S}=\textrm{Tr}_{R}(\rho)$ defined as the unique operator on $\mathcal{F}_{S}$ with 
\begin{equation} \label{eq:FPT4}
\textrm{Tr}_{S}(\rho_{S}O_{S}) = \textrm{Tr}(\rho O_{S})\qquad \forall O_{S}\in \mathcal{A}_{S},
\end{equation} 
where $A_{S}$ is the canonical anti-commutation relation (CAR) algebra of the system. Let $P=(-1)^{N}$ denote the total fermion parity of the bipartite space, with $N$ the total number operator. In~\cite{Amosov_2016}, it is shown that if the modes being traced out are a contiguous set in the mode ordering and the global state $\rho$ is parity symmetric, i.e. $[P,\rho]=0$, then the mode reduced state $\rho_{S}$ coincides with the operator obtained from conventional partial trace methods under a Jordan Wigner transform. 

We will use this when tracing out the reservoir in the definition of $\tau_{\textrm{MF}}$. We thus fix the mode ordering and therefore the definitions of the mode algebras $\mathcal{A}_{S}$ and $\mathcal{A}_{R}$ to be system modes first,

\begin{equation}   \label{eq:FPT5}
\ket{\boldsymbol{n}_{S},\boldsymbol{m}_{R}}= (s_{1}^{\dag})^{n_1}...(s_{L_{S}}^{\dag})^{n_{L_{S}}}(a_{1}^{\dag})^{m_{1}}...(a_{L_R}^{\dag})^{m_{L_{R}}}\ket{\textrm{vac}},
\end{equation}
where $L_{S},L_{R}$ are the number of system and reservoir modes respectively. We will also need to use some properties related to parity in the derivation, which we lay out here. Write $P=P_S P_R$ with $P_S,P_R$ the system and reservoir parities. An operator $O$ is \emph{even} if $POP^{\dagger}=O$ and \emph{odd} if $POP^{\dagger}=-O$; equivalently, even operators contain an even number of mode operators. In Sec.~\ref{sec:sm-kubo}, we use the following property.
Let $O_S\in\mathcal{A}_{S}$ and $O_R\in\mathcal{A}_{R}$. If either $O_S$ or $O_R$ is even, then $[O_S,O_R]=0$. If both are odd, then $\{O_S,O_R\}=0$. This is immediate from the canonical anticommutation relations: moving an odd operator past an odd operator on disjoint modes generates a single minus sign, whereas moving an even one past anything generates none. Fermionic systems follow superselection rules~\cite{Amosov_2016,PhysRevD.1.3267,PhysRev.88.101}, so any physical Hamiltonian is even; $[H_S,O_R]=[H_R,O_S]=0$ for arbitrary $O_{S},O_{R}$, and the same holds for $K_S$ and $K_R$ since number operators are even. This is what allows the $\beta$-interaction picture below to factorise. If $X$ is odd, $X^{\dagger}X$, $XX^{\dagger}$, $X_2^{\dagger}X_1$ and $X_2X_1^{\dagger}$ are all even and commute with reservoir operators; this is what makes the two fermionic swaps in Eqs.~\eqref{eq:SM_derivation_11}--\eqref{eq:SM_derivation_12} cancel.

\subsection{Kubo expansion and the reservoir trace} \label{sec:sm-kubo}

We start from the Kubo expansion~\cite{Thingna2012,Geva2000,MoriMiyashita2008,PhysRevLett.127.250601} 
\begin{equation}
e^{-\beta(K_{0}+\lambda V)} = e^{-\beta K_{0}}\bigg(1-\lambda\int_{0}^{\beta}\!\!d\beta_{1}\,e^{\beta_{1}K_{0}}Ve^{-\beta_{1}K_{0}}+\lambda^{2}\int_{0}^{\beta}\!\!d\beta_{2}\int_{0}^{\beta_{2}}\!\!d\beta_{1}\,e^{\beta_{2}K_{0}}Ve^{-(\beta_{2}-\beta_{1})K_{0}}Ve^{-\beta_{1}K_{0}}+\mathcal{O}(\lambda^{3})\bigg).
\label{eq:SM_derivation_6}
\end{equation}
We define the ``$\beta$-interaction'' picture $V(-i\beta') = e^{\beta'K_{0}}Ve^{-\beta'K_{0}}$. As $K_S$ and $K_R$ commute with reservoir and system operators respectively (Sec.~\ref{app:fermionic partial trace}), we have
\begin{equation}
    V(-i \beta')
    = X^{\dagger}(-i\beta') E(-i\beta')
      + E^{\dagger}(-i\beta') X(-i\beta') ,
\label{eq:SM_derivation_7}
\end{equation}
where
\begin{equation}
    X(-i\beta')=e^{\beta' K_S}Xe^{-\beta'K_S},
    \qquad
    E(-i\beta')=e^{\beta' K_R}Ee^{-\beta'K_R}.
\label{eq:SM_derivation_8}
\end{equation}
With this, Eq.~\eqref{eq:SM_derivation_6} becomes
\begin{equation}
e^{-\beta(K_{0}+\lambda V)} = e^{-\beta K_{0}}\bigg(1-\lambda\int_{0}^{\beta}\!\!d\beta_{1}V(-i\beta_{1})+\lambda^{2}\int_{0}^{\beta}\!\!d\beta_{2}\int_{0}^{\beta_{2}}\!\!d\beta_{1}V(-i\beta_{2})V(-i\beta_{1})+\mathcal{O}(\lambda^{3})\bigg).
\label{eq:SM_derivation_9}
\end{equation}
Introducing the shorthand $X_{j} = X(-i\beta_{j})$, $E_{j} = E(-i\beta_{j})$, the $\mathcal{O}(\lambda^{2})$ term contains four contributions,
\begin{equation}
V(-i\beta_{2})V(-i\beta_{1}) =   X_2^{\dagger} E_2 X_1^{\dagger} E_1
    + X_2^{\dagger} E_2 E_1^{\dagger} X_1
    + E_2^{\dagger} X_2 X_1^{\dagger} E_1
    + E_2^{\dagger} X_2 E_1^{\dagger} X_1 .
\label{eq:SM_derivation_10}
\end{equation}
The reservoir Gibbs state is block diagonal in excitation number, so the first and last terms, which contain $E_2E_1$ and $E_2^{\dagger}E_1^{\dagger}$ respectively and therefore connect different number sectors, vanish on tracing over the reservoir. The remaining two terms factorise into system and reservoir parts. $X_2^{\dagger}X_1$ and $X_2X_1^{\dagger}$ are even-parity system operators, so no net sign is generated when the two system operators are brought together, the two fermionic swaps required in each case cancel,
\begin{align}
    X_2^{\dagger} E_2 E_1^{\dagger} X_1
    &= X_2^{\dagger} X_1\, E_2E_1^{\dagger}, \label{eq:SM_derivation_11}\\
    E_2^{\dagger} X_2 X_1^{\dagger} E_1
    &= X_2X_1^{\dagger}\, E_2^{\dagger} E_1 . \label{eq:SM_derivation_12}
\end{align}
We can now trace over the reservoir. As the reservoir modes are contiguous and all operators involved have even parity, the fermionic partial trace as defined by Eq.~\eqref{eq:FPT4} coincides with the usual partial trace here (Sec.~\ref{app:fermionic partial trace}), leading to the following expression for the unnormalised mean force Gibbs state
\begin{equation}
\tilde{\rho}_{S} = e^{-\beta K_{S}}\bigg(1+\lambda^{2}\int_{0}^{\beta}\!\!d\beta_{2}\int_{0}^{\beta_{2}}\!\!d\beta_{1}\Big[X^{\dagger}_{2}X_{1}\,G_{1}\big(\beta,-i(\beta_{2}-\beta_{1})\big)+X_{2}X^{\dagger}_{1}\,G_{2}\big(\beta,-i(\beta_{2}-\beta_{1})\big)\Big]+\mathcal{O}(\lambda^{4})\bigg),
\label{eq:SM_derivation_13}
\end{equation}
where
\begin{align}
    G_1\big(\beta,-i(\beta_2-\beta_1)\big)
    &= \Trs_R\big[\tau_R E(-i\beta_2)E^{\dagger}(-i\beta_1)\big]
    = \Trs_R\big[\tau_R E\big(-i(\beta_2-\beta_1)\big)E^{\dagger}\big],
    \label{eq:SM_derivation_14}
    \\
    G_2\big(\beta,-i(\beta_2-\beta_1)\big)
    &= \Trs_R\big[\tau_R E^{\dagger}(-i\beta_2)E(-i\beta_1)\big] = \Trs_R\big[\tau_R E^{\dagger}\big(-i(\beta_2-\beta_1)\big)E\big].
    \label{eq:SM_derivation_15}
\end{align}
Dependence on $\beta_2-\beta_1$ alone follows from imaginary-time translation invariance of the equilibrium reservoir state. The $\mathcal{O}(\lambda)$ and $\mathcal{O}(\lambda^{3})$ contributions vanish because they contain odd-parity environment operators, whose reservoir trace against $\tau_R$ is zero. Equations~\eqref{eq:SM_derivation_14}--\eqref{eq:SM_derivation_15} are the imaginary-time analogues of the reservoir temporal correlation functions familiar from open-system dynamics~\cite{10.1116/5.0073853}.

\subsection{Eigenoperator decomposition}

To arrive at a practical expression we decompose $X$ into eigenoperators of $K_S$,
\begin{equation}
X = \sum_{\omega}X_{\omega},
\qquad
X_{\omega} = \sum_{n,m:\,\varepsilon_n-\varepsilon_{m}=\omega}\Pi_{n}X\Pi_{m},
\qquad
[{K_{S}},X_{\omega}] = \omega X_{\omega},
\label{eq:SM_derivation_16}
\end{equation}
where $\Pi_{n}$ projects onto the (possibly degenerate) $\varepsilon_{n}$ eigenspace of $K_S$. The sum over $\omega$ therefore runs over distinct values of $\omega$ rather than over distinct transitions as several transitions may share the same Bohr frequency, so are collected into a single $X_\omega$. This grouping is essential, because separating such transitions would generate spurious $(\omega'-\omega)^{-1}$ terms with vanishing denominators in Eq.~\eqref{eq:SM_derivation_23}.  It follows that
\begin{equation}
X(-i\beta') = e^{\beta' K_{S}}Xe^{-\beta' K_{S}} = \sum_{\omega}e^{\beta'\omega}X_{\omega},
\label{eq:SM_derivation_17}
\end{equation}
and, taking the adjoint of $[K_{S},X_{\omega}] = \omega X_{\omega}$ to obtain $[K_{S},X_{\omega}^{\dagger}] = -\omega X_{\omega}^{\dagger}$,
\begin{equation}
X^{\dagger}(-i\beta') = e^{\beta'K_{S}}X^{\dagger}e^{-\beta' K_{S}} = \sum_{\omega}e^{-\beta'\omega}X_{\omega}^{\dagger}.\label{eq:SM_derivation_18}
\end{equation}
Substituting into Eq.~\eqref{eq:SM_derivation_13},
\begin{align}
\tilde{\rho}_{S} &= e^{-\beta {K_{S}}}\bigg(1+\lambda^{2}\sum_{\omega,\omega'}X^{\dagger}_{\omega}X_{\omega'}\int_{0}^{\beta}\!\!d\beta_{2}\int_{0}^{\beta_{2}}\!\!d\beta_{1}\,e^{-\beta_{2}\omega}e^{\beta_{1}\omega'}G_{1}\big(\beta,-i(\beta_{2}-\beta_{1})\big) \nonumber \\
&\hspace{4.5em}+\lambda^{2}\sum_{\omega,\omega'}X_{\omega}X^{\dagger}_{\omega'}\int_{0}^{\beta}\!\!d\beta_{2}\int_{0}^{\beta_{2}}\!\!d\beta_{1}\,e^{\beta_{2}\omega}e^{-\beta_{1}\omega'}G_{2}\big(\beta,-i(\beta_{2}-\beta_{1})\big)+\mathcal{O}(\lambda^{4})\bigg).\label{eq:SM_derivation_19}
\end{align}
Changing variables and swapping the order of integration gives
\begin{align}
\tilde{\rho}_{S} &= e^{-\beta K_{S}}\bigg(1+\lambda^{2}\sum_{\omega,\omega'}X^{\dagger}_{\omega}X_{\omega'}\int_{0}^{\beta}\!\!d\beta_{1}\int_{\beta_{1}}^{\beta}\!\!d\beta_{2}\,e^{-\beta_{2}(\omega-\omega')}e^{-\beta_{1}\omega'}G_{1}(\beta,-i\beta_{1}) \nonumber \\
&\hspace{4.5em}+\lambda^{2}\sum_{\omega,\omega'}X_{\omega}X^{\dagger}_{\omega'}\int_{0}^{\beta}\!\!d\beta_{1}\int_{\beta_{1}}^{\beta}\!\!d\beta_{2}\,e^{\beta_{2}(\omega-\omega')}e^{\beta_{1}\omega'}G_{2}(\beta,-i\beta_{1})+\mathcal{O}(\lambda^{4})\bigg).\label{eq:SM_derivation_20}
\end{align}
The $\beta_{2}$ integral can now be carried out,
\begin{align}
\tilde{\rho}_{S} &= e^{-\beta K_{S}}\bigg(1+
\lambda^2\sum_{\omega,\omega'}
    \Bigg[\frac{X_\omega^{\dagger} X_{\omega'}}{\omega'-\omega}
    \bigg(
        e^{-\beta(\omega-\omega')}
        \int_0^\beta\!\! d\beta_1\,
        e^{-\beta_1\omega'}G_1(\beta,-i\beta_1)
        -
        \int_0^\beta \!\!d\beta_1\,
        e^{-\beta_1\omega}G_1(\beta,-i\beta_1)
    \bigg)
\nonumber \\
    &\hspace{6em}+
    \frac{X_\omega X_{\omega'}^{\dagger}}{\omega-\omega'}
    \bigg(
        e^{\beta(\omega-\omega')}
        \int_0^\beta \!\!d\beta_1\, e^{\beta_1\omega'}G_2(\beta,-i\beta_1)
        -\int_0^\beta \!\!d\beta_1\,
        e^{\beta_1\omega}G_2(\beta,-i\beta_1)
    \bigg)
    \Bigg]\bigg).\label{eq:SM_derivation_21}
\end{align}
Defining
\begin{equation}
A_{1}(\omega) = \int_{0}^{\beta}\!\!d\tau\, e^{-\tau\omega}G_{1}(\beta,-i\tau),
\qquad
A_{2}(\omega) = \int_{0}^{\beta}\!\!d\tau\, e^{\tau\omega}G_{2}(\beta,-i\tau),
\label{eq:SM_derivation_22}
\end{equation}
this can be written compactly as
\begin{equation}
\tilde{\rho}_{S} = e^{-\beta K_{S}
}\bigg(1+\lambda^{2}\sum_{\omega,\omega'}\bigg[X_{\omega}^{\dagger}X_{\omega'}\frac{e^{-\beta(\omega-\omega')}A_{1}(\omega')-A_{1}(\omega)}{\omega'-\omega}+X_{\omega}X_{\omega'}^{\dagger}\frac{e^{\beta(\omega-\omega')}A_{2}(\omega')-A_{2}(\omega)}{\omega-\omega'}\bigg]\bigg).
\label{eq:SM_derivation_23}
\end{equation}
We now treat the $\omega=\omega'$ and $\omega\neq\omega'$ cases separately. For $\omega=\omega'$, writing $\delta=\omega'-\omega$ and expanding,
\begin{equation}
\lim_{\delta\to 0}\frac{e^{\beta\delta}A_1(\omega+\delta)-A_1(\omega)}{\delta}
= \beta A_1(\omega)+\frac{dA_{1}(\omega)}{d\omega},
\label{eq:SM_derivation_24}
\end{equation}
and
\begin{equation}
-\lim_{\delta\to0}\frac{e^{-\beta\delta}A_2(\omega+\delta)-A_2(\omega)}{\delta} =
\beta A_2(\omega)-\frac{dA_{2}(\omega)}{d\omega}.
\label{eq:SM_derivation_25}
\end{equation}
For the $\omega\neq\omega'$ terms we remove the exponential factors using
\begin{equation}
e^{-\beta K_{S}}X_{\omega}=X_{\omega}e^{-\beta K_{S}}e^{-\beta \omega},
\qquad
e^{-\beta K_{S}}X^{\dagger}_{\omega}=X^{\dagger}_{\omega}e^{-\beta K_{S}}e^{\beta \omega},
\label{eq:SM_derivation_26}
\end{equation}
which follow from $[K_S,X_\omega]=\omega X_\omega$, giving
\begin{equation}
e^{-\beta K_{S}}X_{\omega}X_{\omega'}^{\dagger} = X_{\omega}X_{\omega'}^{\dagger}e^{-\beta K_{S}}e^{-\beta (\omega-\omega')},
\qquad
e^{-\beta K_{S}}X_{\omega}^{\dagger}X_{\omega'} = X_{\omega}^{\dagger}X_{\omega'}e^{-\beta K_{S}}e^{\beta (\omega-\omega')}.
\label{eq:SM_derivation_27}
\end{equation}
Using Eqs.~\eqref{eq:SM_derivation_24}--\eqref{eq:SM_derivation_27}, relabelling $\omega\leftrightarrow\omega'$ in some terms and dividing by $Z_{S}\equiv\Trs(e^{-\beta {K_{S}}})$, the unnormalised state becomes
\begin{align}
\tilde{\rho}_{S} &= \tau_{S}+\lambda^{2}\tilde{\Delta}+\mathcal{O}(\lambda^{4}),\nonumber\\
\tilde{\Delta} &= \sum_{\omega}\bigg[\tau_{S}X_{\omega}^{\dagger}X_{\omega}\bigg(\beta A_{1}(\omega)+\frac{dA_{1}(\omega)}{d\omega}\bigg)+\tau_{S}X_{\omega}X_{\omega}^{\dagger}\bigg(\beta A_{2}(\omega)-\frac{dA_{2}(\omega)}{d\omega}\bigg)\bigg] \nonumber \\
&\quad+ \sum_{\omega\neq\omega'}(\omega'-\omega)^{-1}\bigg[\big(X_{\omega'}X_{\omega}^{\dagger}\tau_{S}+\tau_{S}X_{\omega}X_{\omega'}^{\dagger}\big)A_{2}(\omega)-\big(X_{\omega'}^{\dagger}X_{\omega}\tau_{S}+\tau_{S}X_{\omega}^{\dagger}X_{\omega'}\big)A_{1}(\omega)\bigg].
\label{eq:SM_derivation_28}
\end{align}
To normalise $\tilde{\rho}_{S}$ we require its trace. Taking the trace of the first relation in Eq.~\eqref{eq:SM_derivation_27}, using cyclicity and rearranging gives
\begin{equation}
\big(1-e^{-\beta (\omega-\omega')}\big)\Trs_S\big[e^{-\beta K_{S}}X_{\omega}X_{\omega'}^{\dagger}\big]= 0 .\label{eq:SM_derivation_29}
\end{equation}
Since $\omega\neq\omega'$, this gives $\Trs_S[e^{-\beta K_{S}}X_{\omega}X_{\omega'}^{\dagger}] = 0$, and the same argument applies to the other three off-diagonal terms, so only the diagonal terms of $\tilde{\Delta}$ contribute to the trace. Their coefficients simplify. Differentiating Eq.~\eqref{eq:SM_derivation_22},
\begin{align}
    A_1'(\omega)
    &= -\int_0^\beta\!\!d\tau\,
       \tau\,e^{-\tau\omega}G_1(\beta,-i\tau),
&
    A_2'(\omega)
    &= \int_0^\beta\!\!d\tau\,
       \tau\,e^{\tau\omega}G_2(\beta,-i\tau),\label{eq:SM_derivation_30}
\end{align}
so that
\begin{align}
    D_{1}(\omega)\equiv\beta A_{1}(\omega)+A_1'(\omega)
    &= \int_0^\beta\!\!d\tau\,
       (\beta-\tau)\,e^{-\tau\omega}G_1(\beta,-i\tau),
\label{eq:SM_derivation_31}\\
    D_{2}(\omega)\equiv\beta A_{2}(\omega)-A_2'(\omega)
    &= \int_0^\beta\!\!d\tau\,
       (\beta-\tau)\,e^{\tau\omega}G_2(\beta,-i\tau),
\label{eq:SM_derivation_32}
\end{align}
and hence
\begin{equation}
\tilde{Z}_{S} \equiv \Trs_S\tilde{\rho}_{S} = 1+\lambda^{2}z,
\qquad
z \equiv \Trs_S\tilde{\Delta} = \sum_{\omega}\Big[\langle X_{\omega}^{\dagger}X_{\omega}\rangle_{\tau_S}D_{1}(\omega)+\langle X_{\omega}X_{\omega}^{\dagger}\rangle_{\tau_S}D_{2}(\omega)\Big],
\label{eq:SM_derivation_33}
\end{equation}
where $\langle O\rangle_{\tau_S}\equiv\Trs_S[\tau_S O]$.
We now consider $\tau_{\MF}(\lambda)=\tilde{\rho}_S(\lambda)/\tilde{Z}_S(\lambda)$ as a Taylor expansion following the approach in~\cite{Cerisola2024}, noting that it is even in $\lambda$,
\begin{equation}
\tau_{\MF}(\lambda) = \tau_{\MF}(0)+\tfrac{1}{2}\lambda^{2}\frac{d^{2}\tau_{\MF}}{d\lambda^{2}}(0)+\mathcal{O}(\lambda^{4}).\label{eq:SM_derivation_34}
\end{equation}
Using $\tilde{Z}_S(0)=1$, $\tau_{\MF}(0)=\tau_S$, $\tfrac12\,d^{2}\tilde{\rho}_S/d\lambda^{2}(0)=\tilde{\Delta}$ and $\tfrac12\,d^{2}\tilde{Z}_S/d\lambda^{2}(0)=z$ we find
\begin{equation}
\tau_{\MF}^{(2)} = \tau_{S}+\lambda^{2}\big(\tilde{\Delta}-z\,\tau_{S}\big)+\mathcal{O}(\lambda^{4}).
\label{eq:SM_derivation_35}
\end{equation}
This gives the final result,
\begin{align} 
\tau_{\MF}^{(2)}
&= \tau_{S}+\lambda^{2}\big(\Delta_{\rm diag}+\Delta_{\rm off}\big) \nonumber \\
&=\tau_{S}+\lambda^{2}\sum_{\omega}\bigg[\tau_{S}\Big(X_{\omega}^{\dagger}X_{\omega}-\langle X_{\omega}^{\dagger}X_{\omega}\rangle_{\tau_S}\Big)D_{1}(\omega) + \tau_{S}\Big(X_{\omega}X_{\omega}^{\dagger}-\langle X_{\omega}X_{\omega}^{\dagger}\rangle_{\tau_S}\Big)D_{2}(\omega)\bigg] \nonumber \\
&\quad+\lambda^2\sum_{\omega\neq\omega'}
    (\omega'-\omega)^{-1}\bigg[
   \big(X_{\omega'} X_{\omega}^{\dagger}\tau_{S}+\tau_{S}X_\omega X_{\omega'}^{\dagger}\big) A_2(\omega)
    - \big(X_{\omega'}^{\dagger} X_{\omega}\tau_{S}+\tau_{S}X_\omega^{\dagger} X_{\omega'}\big)A_1(\omega)
    \bigg].\label{eq:SM_derivation_36}
\end{align}
This is Eq.~\eqref{eq:result_2} of the main text.

\subsection{Coherences}
\label{sec:SM-coherences}

Here we show that $\tau_{\MF}^{(2)}$ contains coherences in the eigenbasis of $H_{S}$. Every operator appearing in Eq.~\eqref{eq:SM_derivation_36} conserves $N_S$, so its commutator with $H_S$ equals its commutator with $K_S$ and the shifts of the frequency labels by $\mu$ cancel throughout. We therefore work with $H_S$ and the unshifted Bohr frequencies. For any Bohr frequency $\omega$,
\begin{equation}
[H_S, X_\omega^{\dagger} X_\omega] = [H_S,X_\omega^{\dagger}]X_\omega + X_\omega^{\dagger}[H_S,X_\omega]
= -\omega X_\omega^{\dagger} X_\omega + \omega X_\omega^{\dagger} X_\omega = 0,\label{eq:SM_derivation_37}
\end{equation}
and likewise $[H_S, X_\omega X_\omega^{\dagger}]=0$. Since $[\tau_S,H_S]=0$ it follows that $[H_{S},\Delta_{\rm diag}] = 0$. Now consider $\Delta_{\rm off}$. Using $[H_S,X_{\omega'}X_\omega^{\dagger}] = (\omega'-\omega)X_{\omega'}X_\omega^{\dagger}$ together with the three analogous identities for $X_\omega X_{\omega'}^{\dagger}$, $X_{\omega'}^{\dagger}X_\omega$ and $X_\omega^{\dagger}X_{\omega'}$ gives
\begin{align}
[X_{\omega'}X_\omega^{\dagger}\,\tau_S,\, H_S] &= (\omega-\omega')\,X_{\omega'}X_\omega^{\dagger}\,\tau_S, &
[\tau_S X_\omega X_{\omega'}^{\dagger},\, H_S] &= (\omega'-\omega)\,\tau_S X_\omega X_{\omega'}^{\dagger},\label{eq:SM_derivation_38} \\
[X_{\omega'}^{\dagger} X_\omega\,\tau_S,\, H_S] &= (\omega'-\omega)\,X_{\omega'}^{\dagger} X_\omega\,\tau_S, &
[\tau_S X_\omega^{\dagger} X_{\omega'},\, H_S] &= (\omega-\omega')\,\tau_S X_\omega^{\dagger} X_{\omega'} .\label{eq:SM_derivation_39}
\end{align}
In each case the factor $(\omega'-\omega)$ produced by the commutator cancels the prefactor $(\omega'-\omega)^{-1}$ in Eq.~\eqref{eq:SM_derivation_36}, leaving
\begin{equation}
[\tau_{\MF}^{(2)}, H_S] = \lambda^2 \sum_{\omega\neq\omega'}
\Big\{ A_2(\omega)\big[\tau_S X_\omega X_{\omega'}^{\dagger} - X_{\omega'}X_\omega^{\dagger}\,\tau_S\big]
+ A_1(\omega)\big[\tau_S X_\omega^{\dagger} X_{\omega'} - X_{\omega'}^{\dagger} X_\omega\,\tau_S\big] \Big\}.
\label{eq:SM_derivation_40}
\end{equation}
As in the bosonic case~\cite{PhysRevLett.127.250601}, this is generically non-zero.
\subsection{KMS relation between the two channels}
\label{sec:SM-KMS}

The two correlation functions $G_1$ and $G_2$ are not independent. Since $\tau_R$ is thermal, for $0\le\tau\le\beta$,
\begin{align}
G_{1}(\beta,-i\tau)
&= \frac{1}{Z_R}\Trs_R\Big[e^{-\beta K_R}e^{\tau K_R}Ee^{-\tau K_R}E^{\dagger}\Big]
= \frac{1}{Z_R}\Trs_R\Big[e^{-(\beta-\tau)K_R}E\,e^{-\tau K_R}E^{\dagger}\Big] \nonumber\\
&= \frac{1}{Z_R}\Trs_R\Big[e^{-\tau K_R
}E^{\dagger}\,e^{-(\beta-\tau)K_R}E\Big]
= \frac{1}{Z_R}\Trs_R\Big[e^{-\beta K_R}e^{(\beta-\tau)K_R}E^{\dagger}e^{-(\beta-\tau)K_R}E\Big] \nonumber\\
&= G_{2}\big(\beta,-i(\beta-\tau)\big),
\label{eq:SM_derivation_41}
\end{align}
where the third equality is cyclicity of the trace. This is the imaginary-time KMS condition and requires only that the reservoir be in equilibrium, it holds for fermionic and bosonic reservoirs alike. Substituting Eq.~\eqref{eq:SM_derivation_41} into the definition of $A_1$ and changing variables to $u=\beta-\tau$,
\begin{equation}
A_{1}(\omega) = \int_{0}^{\beta}\!\!d\tau\,e^{-\tau\omega}G_{2}\big(\beta,-i(\beta-\tau)\big)
= e^{-\beta\omega}\int_{0}^{\beta}\!\!du\;e^{u\omega}G_{2}(\beta,-i u)
= e^{-\beta\omega}A_{2}(\omega).
\label{eq:SM_derivation_42}
\end{equation}
Differentiating and using Eq.~\eqref{eq:SM_derivation_31},
\begin{equation}
D_{1}(\omega) = \beta A_1(\omega)+A_1'(\omega)
= \beta e^{-\beta\omega}A_2(\omega) -\beta e^{-\beta\omega}A_2(\omega)+e^{-\beta\omega}A_2'(\omega)
= e^{-\beta\omega}A_{2}'(\omega),
\label{eq:SM_derivation_43}
\end{equation}
Together with Eq.~\eqref{eq:SM_derivation_32}, all four coefficient functions are determined by $A_2(\omega)$ and $A_2'(\omega)$. 
\subsection{Explicit coefficient functions}
\label{sec:SM-coefficients}

We now evaluate $G_{1},G_{2}$ and hence $A_{1,2},D_{1,2}$ for the canonical setup outlined in Sec.~\ref{sec:Setup}. Eqs.~(\ref{eq:SM_derivation_3},\ref{eq:SM_derivation_5}) give $a_{\varepsilon}(-i\tau)=e^{-\tau\xi}a_{\varepsilon}$. Writing $g$ for the thermal occupation, $g=f$ with $f(\xi)=(e^{\beta\xi}+1)^{-1}$ for fermions and $g=n$ with $n(\xi)=(e^{\beta\xi}-1)^{-1}$ for bosons, and $\sigma=+1$ for fermions, $\sigma=-1$ for bosons,
\begin{align}
G_{1}(\beta,-i\tau) &= \int\! d\varepsilon\, J(\varepsilon)\,e^{-\tau\xi}\,\big[1-\sigma\, g(\xi)\big] , \label{eq:SM_derivation_45}\\
G_{2}(\beta,-i\tau) &= \int\! d\varepsilon\, J(\varepsilon)\,e^{\tau\xi}\, g(\xi),
\label{eq:SM_derivation_46}
\end{align}
with $\xi=\varepsilon-\mu$. 
Performing the $\tau$ integrals in Eqs.~\eqref{eq:SM_derivation_22} and~\eqref{eq:SM_derivation_31}--\eqref{eq:SM_derivation_32} with the correlation functions of Eqs.~\eqref{eq:SM_derivation_45}--\eqref{eq:SM_derivation_46}, and writing $x\equiv\xi+\omega=\varepsilon-\mu+\omega$,
\begin{align}
A_{1}(\omega) &= \int\! d\varepsilon\, J(\varepsilon)\big[1-\sigma g({\xi})\big]\,\frac{1-e^{-\beta x}}{x},
\label{eq:SM_derivation_47}\\
A_{2}(\omega) &= \int\! d\varepsilon\, J(\varepsilon)\, g({\xi})\,\frac{e^{\beta x}-1}{x},
\label{eq:SM_derivation_48}\\
D_{1}(\omega) &= \int\! d\varepsilon\, J(\varepsilon)\big[1-\sigma g({\xi})\big]\bigg[\frac{\beta}{x}-\frac{1-e^{-\beta x}}{x^{2}}\bigg],
\label{eq:SM_derivation_49}\\
D_{2}(\omega) &= \int\! d\varepsilon\, J(\varepsilon)\, g({\xi})\bigg[-\frac{\beta}{x}+\frac{e^{\beta x}-1}{x^{2}}\bigg],
\label{eq:SM_derivation_50}
\end{align}
where we used
\begin{equation}
\int_0^{\beta}\!\!d\tau\,(\beta-\tau)\,e^{-\tau x}
= e^{-\beta x}\int_{0}^{\beta}\!\!ds\; s\,e^{sx} = \frac{\beta}{x}-\frac{1-e^{-\beta x}}{x^{2}} .
\label{eq:SM_derivation_51}
\end{equation}
Each bracket is regular at $x=0$: expanding $1-e^{-\beta x}=\beta x-\tfrac12\beta^{2}x^{2}+\tfrac16\beta^{3}x^{3}+\dots$ gives
\begin{equation}
\frac{\beta}{x}-\frac{1-e^{-\beta x}}{x^{2}} = \frac{\beta^{2}}{2}-\frac{\beta^{3}x}{6}+\mathcal{O}(x^{2}).
\label{eq:SM_derivation_52}
\end{equation}

\section{High-temperature limit} \label{sec:SM-highT}

In this section we derive the high-temperature limit, $\beta\to0$, of the second-order fermionic MFG state. Following~\cite{PhysRevLett.127.250601}, we use the product approximation from the Zassenhaus formula~\cite{Suzuki1977},
\begin{equation} \label{eq:SM_highT_1}
e^{-\beta(K_0+\lambda V)} =\frac{1}{2}\Big[e^{-\beta K_0}e^{-\beta\lambda V}+e^{-\beta\lambda V}e^{-\beta K_0}\Big]+\frac{\beta^3}{12}\Big(\lambda\big[K_{0},[K_0,V]\big]+\lambda^2\big[V,[V,K_0]\big]\Big)+O(\beta^4).
\end{equation}
To arrive at Eq.~\eqref{eq:SM_highT_1}, take expansions of $e^{-\beta(K_0+\lambda V)}$, $e^{-\beta K_0}e^{-\beta\lambda V}$, $e^{-\beta\lambda V}e^{-\beta K_0}$ and combine them. Taking the trace and noting that $\textrm{Tr}_{R}[e^{-\beta\lambda V}e^{-\beta K_{R}}] = \textrm{Tr}_{R}[e^{-\beta K_{R}}e^{-\beta\lambda V}]$ gives
\begin{equation} \label{eq:SM_highT_2}
\textrm{Tr}_{R}\big[e^{-\beta(K_0+\lambda V)}\big] = \frac{Z_{R}}{2}\Big\{e^{-\beta K_{S}},\textrm{Tr}_{R}\big[\tau_{R}e^{-\beta\lambda V}\big]\Big\}+\mathcal{O}(\lambda^2\beta^3).
\end{equation}
The $O(\lambda^2\beta^3)$ error comes from the $\lambda^2\big[V,[V,K_0]\big]$ term and $\lambda\big[K_{0},[K_0,V]\big]$ vanishes as it is linear in $V$ so its trace with $\tau_{R}$ gives zero by number conservation. We now expand the reservoir average as
\begin{align} \label{eq:SM_highT_3}
\textrm{Tr}_{R}\big[\tau_{R}e^{-\beta\lambda V}\big] &= 1+\frac{\beta^2\lambda^2}{2}\textrm{Tr}_{R}\big[\tau_{R}V^2\big]+O(\lambda^3\beta^3), \nonumber \\
&= 1+\frac{\beta^2\lambda^2}{2}\big(XX^\dag\langle E^{\dag}E\rangle_{\tau_{R}}+X^\dag X\langle EE^{\dag}\rangle_{\tau_{R}}\big).
\end{align}
 Here we use the fact that odd terms and $EE,E^\dag E^\dag$ terms vanish by number conservation. We have also rearranged $V^{2}$ in the same way as done in Eqs.~(\ref{eq:SM_derivation_10}--\ref{eq:SM_derivation_12}). Using $E=\int d\varepsilon \sqrt{J(\varepsilon)}a_{\varepsilon}$, $\langle a_\varepsilon^{\dag}a_{\varepsilon'}\rangle_{\tau_{R}} = \delta(\varepsilon-\varepsilon')f(\xi)$, and expanding $f(\xi) = \frac{1}{2}-\frac{1}{4}\beta\xi+O(\beta^3\xi^3)$,
 \begin{equation} \label{eq:SM_highT_4}
\langle E E^\dag\rangle_{\tau_{R}} = \int d\varepsilon J(\varepsilon)\big(1-f(\xi)\big) = \frac{1}{2}m_{0}+\frac{1}{4}\beta m_1 +O(\beta^3),
 \end{equation}
\begin{equation} \label{eq:SM_highT_5}
\langle E^\dag E \rangle_{\tau_{R}} = \int d\varepsilon J(\varepsilon)f(\xi) = \frac{1}{2}m_{0}-\frac{1}{4}\beta m_1 +O(\beta^3),
\end{equation}
\begin{equation} \label{eq:SM_highT_6}
m_k \equiv \int d\varepsilon J(\varepsilon)(\varepsilon-\mu)^{k}.
\end{equation}
The $m_1$ terms multiply $\beta^2$ and so will be $O(\beta^3)$ and can be dropped, giving
\begin{equation} \label{eq:SM_highT_7}
\textrm{Tr}_{R}\big[\tau_{R}e^{-\beta\lambda V}\big] = 1+\frac{\lambda^2\beta^2m_0}{4}C+O(\beta^3).
\end{equation}
with $C = \{X,X^\dag\}$. Substituting this into Eq.~\eqref{eq:SM_highT_2} then gives for the un-normalised state
\begin{equation} \label{eq:SM_highT_8}
\tilde{\rho}_S = e^{-\beta K_S}+\frac{\lambda^2 \beta^2 m_0}{8}\{e^{-\beta K_S},C\}+O(\beta^3).
\end{equation}
Using $e^{-\beta K_S}=1+O(\beta)$ which gives $\{e^{-\beta K_{S}},C\} = 2C+O(\beta)$ and dividing by the trace,
\begin{equation} \label{eq:SM_highT_9}
\tau_{\textrm{MF}} = \frac{e^{-\beta K_{S}}+\frac{1}{4}\lambda^2\beta^2 m_0C}{Z_S+\frac{1}{4}\lambda^2\beta^2 m_0 \textrm{Tr}(C)}+O(\beta^3) = \tau_{S}+\frac{1}{4}\lambda^2\beta^2 m_{0}\Big(\frac{C}{Z_{S}}-\tau_{S}\frac{\textrm{Tr}(C)}{Z_{S}}\Big)+O(\beta^3),
\end{equation}
where the denominator has been expanded to first order, with the next term being $O(\beta^4)$. Using $Z_{S} = d_{S}+O(\beta)$, $\tau_{S} = \mathbb{1}/d_{S}+O(\beta)$, we have
\begin{equation} \label{eq:SM_highT_10}
\tau_{\textrm{MF}} = \tau_{S}+\frac{\lambda^2\beta^2 m_0}{4d_{S}}\delta C+O(\beta^3), \quad \delta C \equiv C-\langle C\rangle_0,
\end{equation}
where $\langle C\rangle_0 = \mathbb{1}\textrm{Tr}(C)/d_{S}$. The correction is traceless as required. This also holds for all coupling strengths, provided the temperature well exceeds the coupling energy scale.  If $X$ is linear in system annihilation operators, $X=\sum_iu_is_i$, then $C=\sum_i|u_i|^2\,\mathbb{1}$ and $\delta C=0$, meaning $\tau_{\textrm{MF}}=\tau_{S}+O(\beta^3)$. This is the case for all three models considered in the main text.

\section{Low-temperature limit}
\label{sec:SM-lowT}

In this section we derive the low-temperature limit, $\beta\to\infty$, of the second-order MFG state for a fermionic reservoir. Physically, one expects to replace the system Gibbs state by the ground state of the system and the reservoir Fermi function by a step function. However, care must be taken. 
The coefficient functions in Eq.~\eqref{eq:SM_derivation_36} contain factors that grow exponentially with $\beta$. These are compensated by exponentially small factors in $\tau_S$ and $f$. Finite contributions therefore arise from products of exponentially large and exponentially small factors, which are lost if the small factors are set to zero first. To remedy this, we rewrite the result in terms of coefficient functions that remain bounded as $\beta\to\infty$.

We introduce the particle and hole functions
\begin{equation}
B_-(\omega)\equiv\mathcal{P}\!\!\int\! d\varepsilon\,\frac{J(\varepsilon)\big[1-{\sigma g(\varepsilon-\mu)}\big]}{{\varepsilon-\mu+\omega}},
\qquad
B_+(\omega)\equiv\mathcal{P}\!\!\int\! d\varepsilon\,\frac{J(\varepsilon)\,{g(\varepsilon-\mu)}}{{\varepsilon-\mu+\omega}},
\label{eq:SM_lowT_1}
\end{equation}
where $\mathcal{P}$ denotes the principal value and, as in Sec.~\ref{sec:SM-coefficients}, $g=f,\ \sigma=+1$ for a fermionic reservoir and $g=n,\ \sigma=-1$ for a bosonic one. Since $0\le f\le1$, both remain bounded for fermions as $\beta\to\infty$ at fixed $\omega$.
We proceed to rewrite $A_{1,2},D_{1,2}$ in terms of $B_{-},B_{+}$.
In Eq.~\eqref{eq:SM_derivation_48}, write $g(\xi)e^{\beta x}=g(\xi)e^{\beta\xi}e^{\beta\omega}=[1-\sigma g(\xi)]e^{\beta\omega}$ using the detailed-balance relation $g(\xi)e^{\beta\xi}=1-\sigma g(\xi)$. The integrand of $A_2$ then splits into two terms, each with a simple pole at $\xi=-\omega$ ($x=0)$. The sum is regular, so each term may be taken as a principal value, giving
\begin{equation}
A_2(\omega)=e^{\beta\omega}B_-(\omega)-B_+(\omega),
\qquad
A_1(\omega)=e^{-\beta\omega}A_2(\omega)=B_-(\omega)-e^{-\beta\omega}B_+(\omega),
\label{eq:SM_lowT_2}
\end{equation}
where the second relation uses Eq.~\eqref{eq:SM_derivation_42}. Differentiating Eq.~\eqref{eq:SM_lowT_2}, $A_2'=\beta e^{\beta\omega}B_-+e^{\beta\omega}B_-'-B_+'$. From Eq.~\eqref{eq:SM_derivation_43} and $D_2=\beta A_2-A_2'$,
\begin{equation}
D_1(\omega)=\beta B_-(\omega)+B_-'(\omega)-e^{-\beta\omega}B_+'(\omega),
\qquad
D_2(\omega)=-\beta B_+(\omega)-e^{\beta\omega}B_-'(\omega)+B_+'(\omega).
\label{eq:SM_lowT_3}
\end{equation}
The reason for doing this is that now all exponential dependence on $\beta$ is in explicit factors $e^{\pm\beta\omega}$, multiplying bounded functions. Inserting Eq.~\eqref{eq:SM_lowT_3} into $\tau_SX_\omega^{\dagger}X_\omega D_1+\tau_SX_\omega X_\omega^{\dagger}D_2$ and using Eq.~\eqref{eq:SM_derivation_26}, we get $e^{-\beta\omega}\tau_SX_\omega^{\dagger}X_\omega=X_\omega^{\dagger}\tau_SX_\omega$ and $e^{\beta\omega}\tau_SX_\omega X_\omega^{\dagger}=X_\omega\tau_SX_\omega^{\dagger}$. The diagonal part of $\tilde{\Delta}$ is therefore
\begin{equation}
\sum_\omega\Big\{\beta\big[B_-(\omega)\,\tau_SX_\omega^{\dagger}X_\omega-B_+(\omega)\,\tau_SX_\omega X_\omega^{\dagger}\big]
+B_-'(\omega)\,[\tau_SX_\omega^{\dagger},X_\omega]+B_+'(\omega)\,[\tau_SX_\omega,X_\omega^{\dagger}]\Big\}.
\label{eq:SM_lowT_4}
\end{equation}
Inserting Eq.~\eqref{eq:SM_lowT_2} into $\Delta_{\rm off}$ and using Eq.~\eqref{eq:SM_derivation_26} gives four replacements:
\begin{equation}
e^{\beta\omega}X_{\omega'}X_\omega^{\dagger}\tau_S=X_{\omega'}\tau_SX_\omega^{\dagger},\quad
e^{\beta\omega}\tau_SX_\omega X_{\omega'}^{\dagger}=X_\omega\tau_SX_{\omega'}^{\dagger},\quad
e^{-\beta\omega}X_{\omega'}^{\dagger}X_\omega\tau_S=X_{\omega'}^{\dagger}\tau_SX_\omega,\quad
e^{-\beta\omega}\tau_SX_\omega^{\dagger}X_{\omega'}=X_\omega^{\dagger}\tau_SX_{\omega'}.\label{eq:SM_lowT_5}
\end{equation}
Collecting the coefficients of $B_-(\omega)$ and $B_+(\omega)$,
\begin{equation}
\Delta_{\rm off}=\sum_{\omega\neq\omega'}\frac{1}{\omega'-\omega}\Big\{B_-(\omega)\big([X_{\omega'},\tau_SX_\omega^{\dagger}]+{\rm H.c.}\big)+B_+(\omega)\big([X_{\omega'}^{\dagger},\tau_SX_\omega]+{\rm H.c.}\big)\Big\}.
\label{eq:SM_lowT_6}
\end{equation}
The commutators in Eq.~\eqref{eq:SM_lowT_4} and all of $\Delta_{\rm off}$ are traceless, so
\begin{equation}
z=\Trs\tilde{\Delta}=\beta\sum_\omega\Big\{B_-(\omega)\big\langle X_\omega^{\dagger}X_\omega\big\rangle_{\tau_S}-B_+(\omega)\big\langle X_\omega X_\omega^{\dagger}\big\rangle_{\tau_S}\Big\}.
\label{eq:SM_lowT_7}
\end{equation}
Combining Eqs.~\eqref{eq:SM_lowT_4}--\eqref{eq:SM_lowT_7} and using Eq.~\eqref{eq:SM_derivation_35}, the second-order state in bounded form is
\begin{align}
\tau_{\MF}^{(2)}=\tau_S+\lambda^2\bigg\{&\beta\sum_\omega\Big[B_-(\omega)\,\tau_S\big(X_\omega^{\dagger}X_\omega-\langle X_\omega^{\dagger}X_\omega\rangle_{\tau_S}\big)-B_+(\omega)\,\tau_S\big(X_\omega X_\omega^{\dagger}-\langle X_\omega X_\omega^{\dagger}\rangle_{\tau_S}\big)\Big]\nonumber\\
&+\sum_\omega\Big[B_-'(\omega)\,[\tau_SX_\omega^{\dagger},X_\omega]+B_+'(\omega)\,[\tau_SX_\omega,X_\omega^{\dagger}]\Big]\nonumber\\
&+\sum_{\omega\neq\omega'}\frac{1}{\omega'-\omega}\Big[B_-(\omega)\big([X_{\omega'},\tau_SX_\omega^{\dagger}]+{\rm H.c.}\big)+B_+(\omega)\big([X_{\omega'}^{\dagger},\tau_SX_\omega]+{\rm H.c.}\big)\Big]\bigg\}.
\label{eq:SM_lowT_8}
\end{align}
Equation~\eqref{eq:SM_lowT_8} is an exact rewriting of Eq.~\eqref{eq:SM_derivation_36}, valid at all temperatures. It has the same structure as Eq.~(3) of Ref.~\cite{PhysRevLett.127.250601}, with the single function $D_\beta(\omega)$ of the Hermitian case replaced by two functions $B_\mp(\omega)$ for the particle and hole channels. 

\subsection{\texorpdfstring{The limit $\beta\to\infty$}{The low-temperature limit}}
\label{sec:SM-lowT-limit}

Now we have rewritten Eq.~\eqref{eq:SM_derivation_36} in a bounded form, we can go ahead and take the low temperature limit, where in the following we assume that $K_S$ has a non-degenerate ground state $\ket{g}$. We also restrict to the fermionic case here, although we note that the bounded form Eq.~\eqref{eq:SM_lowT_8} applies to both fermions and bosons. We write $K_S\ket{n}=\varepsilon_n\ket{n}$ and $\Omega_n\equiv\varepsilon_n-\varepsilon_g$, so that $\Omega_n>0$ for $n\neq g$. We also assume $J(\varepsilon)$ is finite and smooth near $\varepsilon=\mu$. The limit is taken in the $\mathcal{O}(\lambda^2)$ coefficient, i.e.\ at $\lambda\to0$ first.

\vspace{0.5cm}

As $\beta\to\infty$, $\tau_S=\ket{g}\bra{g}+\mathcal{O}(e^{-\beta\Omega_1})$, where $\Omega_1$ is the smallest excitation energy, and $f(\varepsilon-\mu)\to\Theta(\mu-\varepsilon)$. Hence
\begin{equation}
B_-(\omega)\to B_-^{0}(\omega)\equiv\mathcal{P}\!\!\int\! d\varepsilon\,{\frac{J(\varepsilon)\,\Theta(\varepsilon-\mu)}{\varepsilon-\mu+\omega}},
\qquad
B_+(\omega)\to B_+^{0}(\omega)\equiv\mathcal{P}\!\!\int\! d\varepsilon\,{\frac{J(\varepsilon)\,\Theta(\mu-\varepsilon)}{\varepsilon-\mu+\omega}}.
\label{eq:SM_lowT_9}
\end{equation}
Because $B_\pm$ and $B_\pm'$ stay bounded, every term of Eq.~\eqref{eq:SM_lowT_8} without an explicit factor of $\beta$ may be evaluated by these substitutions. Because $[K_S,X_\omega^{\dagger}X_\omega]=0$ and $\ket{g}$ is non-degenerate, $X_\omega^{\dagger}X_\omega\ket{g}\propto\ket{g}$. This gives
\begin{equation}
\ket{g}\bra{g}\Big(X_\omega^{\dagger}X_\omega-\bra{g}X_\omega^{\dagger}X_\omega\ket{g}\Big)=0,\label{eq:SM_lowT_10}
\end{equation}
and likewise for $X_\omega X_\omega^{\dagger}$. The first line of Eq.~\eqref{eq:SM_lowT_8} therefore vanishes at $T=0$. With $\tau_S\to\ket{g}\bra{g}$, each remaining term is non-zero only for particular signs of $\omega$. The $B_-$ and $B_-'$ terms contain $\bra{g}X_\omega^{\dagger}$ or $X_\omega\ket{g}$, so they are non-zero only if $X_\omega\ket{g}\neq0$. That requires $\omega=\Omega_m>0$ for some state $m$. The pole of $B_-^0(\Omega_m)$ at $\varepsilon=\mu-\Omega_m<\mu$ then lies outside the support of $\Theta(\varepsilon-\mu)$. Similarly, the $B_+$ and $B_+'$ terms require $X_\omega^{\dagger}\ket{g}\neq0$, i.e.\ $\omega=-\Omega_m<0$. The pole at $\varepsilon=\mu+\Omega_m>\mu$ lies outside the support of $\Theta(\mu-\varepsilon)$. Every integral that survives is therefore an ordinary, non-singular integral, and no principal value is needed.
With $\tau_S\to\ket{g}\bra{g}$ and $B_\pm\to B_\pm^0$,
\begin{align}
\lim_{\beta\to\infty}\tau_{\MF}^{(2)}=\ket{g}\bra{g}+\lambda^2\bigg\{&\sum_\omega\Big[B_-^{0\prime}(\omega)\,\big[\ket{g}\bra{g}X_\omega^{\dagger},X_\omega\big]+B_+^{0\prime}(\omega)\,\big[\ket{g}\bra{g}X_\omega,X_\omega^{\dagger}\big]\Big]\nonumber\\
&+\sum_{\omega\neq\omega'}\frac{1}{\omega'-\omega}\Big[B_-^0(\omega)\big([X_{\omega'},\ket{g}\bra{g}X_\omega^{\dagger}]+{\rm H.c.}\big)+B_+^0(\omega)\big([X_{\omega'}^{\dagger},\ket{g}\bra{g}X_\omega]+{\rm H.c.}\big)\Big]\bigg\}.
\label{eq:SM_lowT_11}
\end{align}
At $T=0$ the reservoir is a filled Fermi sea, and Eq.~\eqref{eq:SM_lowT_11} is the reduced state of the ground state of $H-\mu N$ to second order, namely $\ket{g}$ dressed by virtual transfer of a particle into the empty states above $\mu$ $(B_{-}^{(0)})$ or of a reservoir particle from below $\mu$ into the system $(B_{+}^{(0)})$. 
One key feature of Eq.~\eqref{eq:SM_lowT_11} is that coherences persist in this limit. Taking the commutator with $H_S$ gives
\begin{align}
\lim_{\beta\to\infty}\big[\tau_{\MF}^{(2)},H_S\big]
= \lambda^{2}\!\!\sum_{\omega\neq\omega'}\!\Big\{&B_-^{0}(\omega)\big(X_\omega\ket{g}\bra{g}X_{\omega'}^{\dagger}+\ket{g}\bra{g}X_\omega^{\dagger}X_{\omega'}\big)\nonumber\\
&+B_+^{0}(\omega)\big(X_{\omega'}^{\dagger}\ket{g}\bra{g}X_\omega+X_{\omega'}X_\omega^{\dagger}\ket{g}\bra{g}\big)\Big\}-{\rm H.c.},
\label{eq:SM_lowT_12}
\end{align}
which is generically non-zero.

\section{Hamiltonian of mean force}
\label{sec:HMF}

The mean force Gibbs state is often expressed through the Hamiltonian of mean
force (HMF)~\cite{RevModPhys.92.041002,Miller2018}, the temperature-dependent
operator for which $\tau_{\MF}$ takes a Gibbs form. Here we derive the HMF to
second order in $\lambda$ for the exchange coupling of Eq.~\eqref{eq:SM_derivation_2}, following the strategy of Ref.~\cite{TimofeevTrushechkin2022}.

We define the grand canonical HMF through
\begin{equation}
  e^{-\beta K_{\MF}} \equiv \frac{1}{Z_R}\,
  \Trs_R\, e^{-\beta(H-\mu N)}.
  \label{eq:SM_HMF_1}
\end{equation}
Dividing by the bare reservoir partition function means $K_{\textrm{MF}}\to K_{S}$ for $V = 0$~\cite{RevModPhys.92.041002}. Because $[H,N]=0$, and the partial
trace is cyclic for the even reservoir operator $N_R$, we have $ [N_S,\Trs_R\,e^{-\beta(H-\mu N)}] = \Trs_R\,[N,e^{-\beta(H-\mu N)}] = 0.
$ Hence $[K_{\MF},N_S]=0$, and $H_{\MF}\equiv K_{\MF}+\mu N_S$ is well
defined, with $\tau_{\MF}=e^{-\beta(H_{\MF}-\mu N_S)}/Z_{\MF}$. 
We seek $K_{\MF}^{(2)}=K_S+\lambda^2K^{(2)}+\mathcal{O}(\lambda^4)$. To do this, we will compare the Kubo expansion of Eq.~\eqref{eq:SM_HMF_1} in terms of this expansion in $\lambda$ with the corresponding Kubo expansion of the un-normalised MFG state, Eq.~\eqref{eq:SM_derivation_23}.

\vspace{0.5cm}

The Kubo
expansion with $K_S$ as reference gives
\begin{equation}
  e^{-\beta(K_S+\lambda^2K^{(2)})}
  = e^{-\beta K_S}\Big[1-\lambda^2\!\int_0^\beta\! d\beta'\,
    e^{\beta'K_S}K^{(2)}e^{-\beta'K_S}+\mathcal{O}(\lambda^4)\Big].
  \label{eq:SM_HMF_2}
\end{equation}
Decomposing $K^{(2)}$ into frequency components $K^{(2)}=\sum_{\nu}K_{\nu}^{(2)}$ where $[K_{S},K_{\nu}^{(2)}] = \nu K_{\nu}^{(2)}$ and using Eq.~\eqref{eq:SM_derivation_17} we can evaluate this integral, giving
\begin{equation}
  e^{-\beta(K_S+\lambda^2K^{(2)})}
  = e^{-\beta K_S}\Big[1-\lambda^2\big[\beta K_{\nu=0}^{(2)}+\sum_{\nu\neq0}K_{\nu}^{(2)}\frac{(e^{\beta\nu}-1)}{\nu}\big]+\mathcal{O}(\lambda^4)\Big],\label{eq:SM_HMF_3}
\end{equation}
where the $\nu=0$ case has been separated for ease of comparison with Eq.~\eqref{eq:SM_derivation_23}, which we will now consider. Note that the operators in Eq.~\eqref{eq:SM_derivation_23} are already eigenoperators of $K_{S}$: $X^\dagger_\omega X_{\omega'}$ has eigenvalue $\nu = \omega'-\omega$ and
$X_\omega X^\dagger_{\omega'}$ has eigenvalue $\nu = \omega-\omega'$. Equating the off-diagonal terms with $\nu=\omega-\omega'$
\begin{align}
K_{\nu}^{(2)} &= -X_{\omega'}^{\dag}X_{\omega}\frac{\omega-\omega'}{e^{\beta(\omega-\omega')}-1}\frac{e^{-\beta(\omega'-\omega)}A_{1}(\omega)-A_{1}(\omega')}{\omega-\omega'}-X_{\omega}X_{\omega'}^{\dag}\frac{\omega-\omega'}{e^{\beta(\omega-\omega')}-1}\frac{e^{\beta(\omega-\omega')}A_{2}(\omega')-A_{2}(\omega)}{\omega-\omega'}, \nonumber \\
&= -X_{\omega'}^{\dag}X_{\omega}\frac{e^{-\beta(\omega'-\omega)}A_{1}(\omega)-A_{1}(\omega')}{e^{\beta(\omega-\omega')}-1}-X_{\omega}X_{\omega'}^{\dag}\frac{e^{\beta(\omega-\omega')}A_{2}(\omega')-A_{2}(\omega)}{e^{\beta(\omega-\omega')}-1}, \nonumber 
\\
&= -X_{\omega'}^{\dag}X_{\omega}\frac{A_{2}(\omega)-A_{2}(\omega')}{e^{\beta\omega}-e^{\beta\omega'}} - X_{\omega}X_{\omega'}^{\dag}\frac{A_{1}(\omega')-A_{1}(\omega)}{e^{-\beta\omega'}-e^{-\beta\omega}}.\label{eq:SM_HMF_4}
\end{align}
where in the last line the KMS relation $A_{1}(\omega)=e^{-\beta\omega}A_{2}(\omega)$ was used and both numerators and denominators were divided by $e^{-\beta\omega'}$ and $e^{\beta\omega}$ for the first and second terms respectively. For the diagonal component, we have
\begin{equation}
K_{\nu=0}^{(2)} = -\frac{1}{\beta}\sum_{\omega}\Big[D_{1}(\omega)X_{\omega}^{\dag}X_{\omega}+D_{2}(\omega)X_{\omega}X_{\omega}^{\dag}\Big].\label{eq:SM_HMF_5}
\end{equation}
Putting this together, the HMF to second order is given by
\begin{align}
  H_{\MF}^{(2)} &= H_S+\lambda^2K^{(2)}+\mathcal{O}(\lambda^4),\nonumber\\
  K^{(2)} &= -\frac{1}{\beta}\sum_\omega\Big[D_1(\omega)\,X^\dagger_\omega X_\omega
      +D_2(\omega)\,X_\omega X^\dagger_\omega\Big]
  -\sum_{\omega\neq\omega'}\bigg[
      \frac{A_2(\omega')-A_2(\omega)}{e^{\beta\omega'}-e^{\beta\omega}}\,X^\dagger_\omega X_{\omega'}
     +\frac{A_1(\omega')-A_1(\omega)}{e^{-\beta\omega'}-e^{-\beta\omega}}\,X_\omega X^\dagger_{\omega'}\bigg],
  \label{eq:SM_HMF_6}
\end{align}
where $H_{\MF}^{(2)}=K_{\MF}^{(2)}+\mu N_S=H_S+\lambda^2K^{(2)}$ because $K^{(2)}$ conserves
$N_S$. However, this expression is more restrictive than Eq.\eqref{eq:SM_derivation_36}. Consider a non-degenerate eigenstate $\ket{n}$ of $H_S$ with energy $E_n$, particle number $N_n$ and bare population $p_n = e^{-\beta(E_n-\mu N_n)}/Z_S$, and suppose for simplicity that $\tau_{\rm MF}^{(2)}$ is diagonal in this basis, as for the RLM and the SIAM. The population of $\ket{n}$ in $\tau_{\rm MF}^{(2)}$, Eq.~\eqref{eq:SM_derivation_36}, is then
\begin{equation}
\bra{n}\tau_{\rm MF}^{(2)}\ket{n} = p_n + \lambda^2\Delta_{nn}= p_n(1+x_n), \qquad x_n = \frac{\lambda^2\Delta_{nn}}{p_{n}},
\end{equation}
where $\Delta_{nn} = \bra{n}\Delta_{\rm diag}\ket{n}$.
By definition, $H_{\rm MF}-\mu N_S = -T\ln\tau_{\rm MF} + \textrm{const}$. Applying this to $\tau_{\rm MF}^{(2)}$, and using $\bra{n}\ln\tau_{\rm MF}^{(2)}\ket{n} = \ln\bra{n}\tau_{\rm MF}^{(2)}\ket{n}$ since $\tau_{\rm MF}^{(2)}$ is diagonal, the Hamiltonian of mean force that reproduces $\tau_{\rm MF}^{(2)}$ exactly has diagonal elements
\begin{equation}
-T\ln\bra{n}\tau_{\rm MF}^{(2)}\ket{n} = E_n - \mu N_n - T\ln(1+x_n) + \textrm{const}.
\end{equation}
$H^{(2)}_{\rm MF} = H_S + \lambda^2K^{(2)}$ of Eq.~\eqref{eq:SM_HMF_6} has diagonal elements
\begin{equation}
\bra{n}H^{(2)}_{\rm MF}-\mu N_S\ket{n} = E_n - \mu N_n - Tx_n + \textrm{const}.
\end{equation}
Here const denotes terms independent of $n$, which only affect the normalisation. The grand canonical Gibbs state of $H^{(2)}_{\rm MF}$ therefore reproduces $\tau_{\rm MF}^{(2)}$ only if $\ln(1+x_n)\simeq x_n$, i.e.\ $|x_n|\ll1$. At low temperature this fails for excited states, their bare population $p_{n}$ vanishes exponentially while the mean force correction $\Delta_{nn}$ remains finite as $T\to0$ (See Sec.~\ref{sec:SM-lowT}), so $x_n$ grows exponentially. 

\section{Applications}

\subsection{Resonant level model}
\label{sec:SM-RLM}

The simplest application is the resonant level model (RLM), a single fermionic mode coupled to a continuous non-interacting reservoir,
\begin{equation}
H = \varepsilon_{s} s^{\dagger}s+\int\! d\varepsilon\,\varepsilon\, a^{\dagger}_\varepsilon a_\varepsilon+\lambda\!\int\!\! d\varepsilon\,\sqrt{J(\varepsilon)}\big(s^{\dagger}a_\varepsilon+a^{\dagger}_\varepsilon s\big),
\label{eq:SM_applications_1}
\end{equation}
so that $X = s$ and $E=\int\sqrt{J(\varepsilon)}a_\varepsilon\, d\varepsilon$. Here $X$ is already an eigenoperator,
\begin{equation}
[{K_{S}},s]=-{(\varepsilon_{s}-\mu)} s \quad \implies \quad X = X_{\omega_{0}}=s, \qquad \omega_{0}={\mu-\varepsilon_{s}},\label{eq:SM_applications_2}
\end{equation}
this being the only non-zero Bohr-frequency component, so there are no off-diagonal contributions. With $X_{\omega_{0}}^{\dagger}X_{\omega_{0}}=s^{\dagger}s\equiv n$, $X_{\omega_{0}}X_{\omega_{0}}^{\dagger}=ss^{\dagger}=1-n$ and $\tau_{S}=\mathrm{diag}\big(1-f_s,f_s\big)$ with $f_s\equiv f(\varepsilon_s-\mu)$, Eq.~\eqref{eq:SM_derivation_36} gives
\begin{equation}
\tau_{\MF}^{(2)} = \tau_{S}+\lambda^{2}\Big[\tau_{S}\big(n-\Trs[\tau_{S}n]\big)D_{1}(\omega_{0})+\tau_{S}\big((1-n)-\Trs[\tau_{S}(1-n)]\big)D_{2}(\omega_{0})\Big].\label{eq:SM_applications_3}
\end{equation}
Since $n-\langle n\rangle_{\tau_S}$ and $(1-n)-\langle 1-n\rangle_{\tau_S}$ are negatives of one another, there is a single correction to the occupation,
\begin{equation} \label{eq:SM_applications_4}
\langle n\rangle_{\MF}^{(2)} = {f_s}+\lambda^{2}{f_s\big[1-f_s\big]}\big[D_{1}(\omega_{0})-D_{2}(\omega_{0})\big].
\end{equation}
To evaluate $D_1 - D_2$, it is easier to work with the bounded form Eq.~\eqref{eq:SM_lowT_3}. With $\omega_0=\mu-\varepsilon_s$ the denominators in Eq.~\eqref{eq:SM_lowT_1} become $\varepsilon-\varepsilon_s$, so
\begin{equation}
B_-(\omega_0)+B_+(\omega_0)=\mathcal{P}\!\!\int\! d\varepsilon\,\frac{J(\varepsilon)}{\varepsilon-\varepsilon_s},
\qquad
f_sB_-'(\omega_0)-(1-f_s)B_+'(\omega_0)=-\mathcal{P}\int\! d\varepsilon\,J(\varepsilon)\,\frac{f_s-f(\varepsilon-\mu)}{(\varepsilon-\varepsilon_s)^2},
\label{eq:SM_applications_7}
\end{equation}
where we used $f_s[1-f(\varepsilon-\mu)]-(1-f_s)f(\varepsilon-\mu)=f_s-f(\varepsilon-\mu)$. The correction to the occupation is therefore
\begin{equation}
\delta\langle n\rangle
= \lambda^{2}\!\int\! d\varepsilon\, J(\varepsilon)\bigg[\frac{\beta f_s(1-f_s)}{\varepsilon-\varepsilon_{s}}-\frac{f_s-f(\varepsilon-\mu)}{(\varepsilon-\varepsilon_{s})^{2}}\bigg]
\label{eq:SM_applications_8}
\end{equation}
The integrand is regular at $\varepsilon=\varepsilon_s$: expanding $f(\varepsilon-\mu)=f_s+f'(\varepsilon_s-\mu)(\varepsilon-\varepsilon_s)+\tfrac12 f''(\varepsilon_s-\mu)(\varepsilon-\varepsilon_s)^2+\dots$ and using $\beta f_s(1-f_s)=-f'(\varepsilon_s-\mu)$, the two divergent pieces cancel and the integrand tends to $\tfrac12 J(\varepsilon_s)f''(\varepsilon_s-\mu)$. We define the level shift
\begin{equation}
\Lambda(\varepsilon_s)\equiv \lambda^{2}\,\mathcal{P}\!\!\int\! d\varepsilon\,\frac{J(\varepsilon)}{\varepsilon_{s}-\varepsilon},
\label{eq:SM_applications_9}
\end{equation}
which is the real part of the retarded self-energy of the level evaluated at the bare pole, familiar from the Green's function treatment of the RLM~\cite{Ochoa2016,Dou2018}. Then
\begin{equation}
\delta\langle n\rangle = \frac{\partial f_s}{\partial\varepsilon_s}\,\Lambda(\varepsilon_s)
\;-\;\lambda^{2}\,\mathcal{P}\!\!\int\! d\varepsilon\, J(\varepsilon)\,\frac{f_s-f(\varepsilon-\mu)}{(\varepsilon-\varepsilon_{s})^{2}} .
\label{eq:SM_applications_10}
\end{equation}
The first term is the naive expectation, the hybridisation renormalises the level by $\Lambda$, and the occupation responds with the Fermi-function susceptibility. The second term is not captured by any level shift, it represents  the broadening of the level, sampling $f$ away from $\varepsilon_s$. We can recast this in terms of the Hamiltonian of mean force, where Eq.~\eqref{eq:SM_HMF_6} reduces to 
\begin{equation}
H_{\textrm{MF}}= \tilde{\varepsilon}_{s}n -\frac{\lambda^2}{\beta} D_2(\omega_0), \qquad \tilde{\varepsilon}_{s} = \varepsilon_{s}-\lambda^{2}\beta^{-1}\big[D_{1}(\omega_{0})-D_{2}(\omega_{0})\big].\label{eq:SM_applications_11}
\end{equation}

\subsection{Two independent modes coupled to a shared reservoir}
\label{sec:SM-two-mode}

We now consider two modes that do not interact directly but share a reservoir. This is the minimal setting in which the reservoir can mediate correlations, and it is the exchange-coupled analogue of the multi-qubit spin--boson problem studied in Refs.~\cite{Chiu2022,LewisWilliamson2026}. Take
\begin{equation}
H_S = \varepsilon_1 n_1 + \varepsilon_2 n_2, \qquad n_i = s_i^{\dagger} s_i,\label{eq:SM_applications_13}
\end{equation}
with no direct hopping between the modes, coupled to a single reservoir through the symmetric combination
\begin{equation}
X = s_1+s_2 .\label{eq:SM_applications_14}
\end{equation}
The reservoir Hamiltonian and $E$ are as in Sec.~\ref{sec:SM-RLM}. Since $[K_S,s_i]=-(\varepsilon_i-\mu) s_i$, each mode is already a Bohr-frequency eigenoperator:
\begin{equation}
X_{\omega_1}=s_1,\quad \omega_1=\mu-\varepsilon_1,\qquad\qquad X_{\omega_2}=s_2,\quad \omega_2=\mu-\varepsilon_2 .\label{eq:SM_applications_15}
\end{equation}
No basis rotation is required, so $X=X_{\omega_1}+X_{\omega_2}$ with exactly two Bohr frequencies. The diagonal part is
\begin{align}
\Delta_{\rm diag} = {}& \tau_S\big(n_1 - \langle n_1\rangle_{\tau_S}\big)D_1(\omega_1) + \tau_S\big((1-n_1) - \langle 1-n_1\rangle_{\tau_S}\big)D_2(\omega_1) \nonumber\\
& + \tau_S\big(n_2 - \langle n_2\rangle_{\tau_S}\big)D_1(\omega_2) + \tau_S\big((1-n_2) - \langle 1-n_2\rangle_{\tau_S}\big)D_2(\omega_2),\label{eq:SM_applications_16}
\end{align}
and the off-diagonal part is
\begin{align}
\Delta_{\rm off} &= \frac{1}{\varepsilon_1-\varepsilon_2}\Big[\big(s_2 s_1^{\dagger}\,\tau_S + \tau_S s_1 s_2^{\dagger}\big)A_2(\omega_1) - \big(s_2^{\dagger} s_1\,\tau_S + \tau_S s_1^{\dagger} s_2\big)A_1(\omega_1) \nonumber\\
&\hspace{4.5em}-\big(s_1 s_2^{\dagger}\,\tau_S + \tau_S s_2 s_1^{\dagger}\big)A_2(\omega_2) + \big(s_1^{\dagger} s_2\,\tau_S + \tau_S s_2^{\dagger} s_1\big)A_1(\omega_2)\Big].
\label{eq:SM_applications_17}
\end{align}
The two modes are independent under $\tau_S$, which factorises into single-mode fermionic factors,
\begin{equation}
\tau_S = \prod_{i=1,2}\Big[(1-f_i)\big(1-n_i\big)+f_i\,n_i\Big],
\qquad f_i\equiv f(\varepsilon_i-\mu).\label{eq:SM_applications_18}
\end{equation}
Labelling the Fock states by their occupations $\nu_i\in\{0,1\}$, the projector onto $\ket{\nu_1\nu_2}$ is $P_{\nu}=\prod_i\big[\nu_i n_i+(1-\nu_i)(1-n_i)\big]$, and the bare populations are $p^{(0)}_{\nu}=\langle P_{\nu}\rangle_{\tau_S}=\prod_i f_i^{\,\nu_i}(1-f_i)^{1-\nu_i}$. Using $(1-n_i)-\langle 1-n_i\rangle_{\tau_S}=-\big(n_i-\langle n_i\rangle_{\tau_S}\big)$ and defining
$\mathcal{D}_i \equiv D_2(\omega_i)-D_1(\omega_i)$ we have $\Delta_{\rm diag}=-\sum_i\tau_S\big(n_i-f_i\big)\mathcal{D}_i$. Since $P_{\nu}$, $\tau_S$ and $n_i$ are all functions of the number operators they commute, so that $\Trs_S\big[P_{\nu}\,\tau_S(n_i-f_i)\big]=p^{(0)}_{\nu}(\nu_i-f_i)$, and the second-order population corrections $\delta p_{\nu}$, defined by $p_{\nu}=p^{(0)}_{\nu}+\delta p_{\nu}$, are

\begin{align}
\delta p_{00} &= \lambda^2(1-f_1)(1-f_2)\big[f_1\,\mathcal{D}_1 + f_2\,\mathcal{D}_2\big],\label{eq:SM_applications_19} \\
\delta p_{01} &= \lambda^2f_2(1-f_1)\big[f_1\,\mathcal{D}_1 - (1-f_2)\,\mathcal{D}_2\big],\label{eq:SM_applications_20} \\
\delta p_{10} &= \lambda^2f_1(1-f_2)\big[f_2\,\mathcal{D}_2 - (1-f_1)\,\mathcal{D}_1\big],\label{eq:SM_applications_21} \\
\delta p_{11} &= -\lambda^2f_1 f_2\big[(1-f_1)\,\mathcal{D}_1+(1-f_2)\,\mathcal{D}_2\big],\label{eq:SM_applications_22}
\end{align}
which sum to zero as required by trace preservation. Summing $\delta p_{10}+\delta p_{11}$ to obtain the correction to $\langle n_1\rangle$, the terms proportional to $\mathcal{D}_2$ cancel, leaving
\begin{equation}
\delta\langle n_1\rangle = -\lambda^2 f_1(1-f_1)\,\mathcal{D}_1 = \lambda^2 f_1(1-f_1)\big[D_1(\omega_1)-D_2(\omega_1)\big],\label{eq:SM_applications_23}
\end{equation}
with no dependence on $\mathcal{D}_2$, and by the $1\leftrightarrow2$ symmetry of the construction
\begin{equation}
\langle n_1\rangle = f_1 + \lambda^2f_1(1-f_1)\big[D_1(\omega_1)-D_2(\omega_1)\big], \qquad \langle n_2\rangle = f_2 + \lambda^2f_2(1-f_2)\big[D_1(\omega_2)-D_2(\omega_2)\big].
\label{eq:SM_applications_24}
\end{equation}
This is exactly the single-mode RLM formula, Eq.~\eqref{eq:SM_applications_4}, applied independently to each mode. At this order each mode's population is completely blind to the other mode's presence, and to the fact that the reservoir is shared. The modes are not, however, blind to one another in every respect, as they would be in the bare Gibbs state, they acquire a coherence, $\big\langle s_2^{\dagger}s_1\big\rangle_{\tau_{\MF}^{(2)}}$. Now only $\Delta_{\textrm{off}}$ contributes and only the terms of Eq.~\eqref{eq:SM_applications_17} containing $s_2s_1^{\dagger}$ or $s_1^{\dagger}s_2$ survive, the others contain $s_is_i$ or $s_i^{\dagger}s_i^{\dagger}$ and vanish. The surviving terms reduce by anticommutation to
\begin{align}
s_2^{\dagger}s_1\,s_2s_1^{\dagger} &= -n_2(1-n_1), &
s_2^{\dagger}s_1\,s_1^{\dagger}s_2 &= (1-n_1)n_2, \nonumber\\
s_1^{\dagger}s_2\,s_2^{\dagger}s_1 &= (1-n_2)n_1, &
s_2s_1^{\dagger}\,s_2^{\dagger}s_1 &= -(1-n_2)n_1 ,
\label{eq:SM_applications_25}
\end{align}
with $\tau_S$ averages $-f_2(1-f_1)$, $f_2(1-f_1)$, $f_1(1-f_2)$ and $-f_1(1-f_2)$. Collecting them,
\begin{equation}
\big\langle s_2^{\dagger}s_1\big\rangle_{\tau_{\MF}^{(2)}} = \lambda^{2}\,\frac{f_2(1-f_1)\big[A_1(\omega_2)-A_2(\omega_1)\big] + f_1(1-f_2)\big[A_2(\omega_2)-A_1(\omega_1)\big]}{\varepsilon_1-\varepsilon_2}.
\label{eq:SM_applications_26}
\end{equation}
This can be simplified using the KMS relation, which gives $f_i\,A_1(\omega_i)=(1-f_i)\,A_2(\omega_i)$. Grouping the four terms of Eq.~\eqref{eq:SM_applications_26} by frequency gives,
\begin{align}
  f_2(1-f_1)\,A_1(\omega_2)+f_1(1-f_2)\,A_2(\omega_2)
    &=\big[(1-f_1)+f_1\big](1-f_2)\,A_2(\omega_2)=(1-f_2)\,A_2(\omega_2),\nonumber\\
  f_2(1-f_1)\,A_2(\omega_1)+f_1(1-f_2)\,A_1(\omega_1)
    &=\big[f_2+(1-f_2)\big](1-f_1)\,A_2(\omega_1)=(1-f_1)\,A_2(\omega_1),
  \label{eq:SM_applications_27}
\end{align}
Using the definition of $A_2(\omega)$ and rearranging, the coherence can be written as
\begin{align}
  \big\langle s_2^\dagger s_1\big\rangle_{\tau^{(2)}_{\mathrm{MF}}}
  &=\lambda^2\,\frac{\Phi(\varepsilon_1)-\Phi(\varepsilon_2)}{\varepsilon_1-\varepsilon_2}
  \label{eq:SM_applications_28}, \quad 
  \Phi(\varepsilon)\equiv\int d\varepsilon'\,J(\varepsilon')\,
  \frac{f(\varepsilon'-\mu)-f(\varepsilon-\mu)}{\varepsilon'-\varepsilon}.
\end{align}
The integrand of $\Phi$ is regular at $\varepsilon'=\varepsilon$, tending to $J(\varepsilon)f'(\varepsilon-\mu)$ at $\varepsilon'\to\varepsilon$. The presence of coherence can be understood through the Hamiltonian of mean force. Evaluating Eq~\eqref{eq:SM_HMF_6} gives
\begin{align}
  H_{\mathrm{MF}}=\sum_{i=1,2}\tilde\varepsilon_i\,n_i
  +g\,\big(s_1^\dagger s_2+s_2^\dagger s_1\big) - \frac{\lambda^2}{\beta}\big[D_2(\omega_1) + D_2(\omega_2)\big],\qquad
  g=\lambda^2\,\frac{\Phi(\varepsilon_1)-\Phi(\varepsilon_2)}{f_1-f_2} .
  \label{eq:SM_applications_29}
\end{align}

with $\tilde\varepsilon_i$ the renormalised RLM level of mode $i$. The MFG state is thus the Gibbs
state of two modes coupled by a reservoir-mediated hopping, and the coherence of
Eq.~\eqref{eq:SM_applications_28} reflects the rotation of their normal modes away from
$s_1,s_2$. Diagonalising Eq.\eqref{eq:SM_applications_29} gives normal modes $b_{\pm}$ defined as $s_{1}=\textrm{cos}\theta\; b_{+}-\textrm{sin}\theta \;b_{-}$, $s_{2}=\textrm{sin}\theta\; b_{+}+\textrm{cos}\theta \;b_{-}$ with eigenvalues $E_{\pm}$ and angle defined by $\textrm{tan}2\theta=2g/(\tilde{\varepsilon_{1}}-\tilde{\varepsilon_{2}})$. The gibbs state for $H_{\textrm{MF}}$ has thermal occupations $\langle b_{k}^{\dag}b_{l}\rangle = \delta_{kl}f(E_{k}-\mu)$. Using $\textrm{sin}2\theta = 2g/(E_{+}-E_{-})$ this gives

\begin{equation}
\big\langle s_2^\dagger s_1\big\rangle = g\frac{f(E_{+}-\mu)-f(E_{-}-\mu)}{E_{+}-E_{-}} \approx g\frac{f_1-f_2}{\varepsilon_{1}-\varepsilon_{2}}\label{eq:SM_applications_30},
\end{equation}
where the approximate equality follows from the fact that $g=\mathcal{O}(\lambda^2)$, so the divided difference is only needed at zeroth order, where $E_{\pm}=\varepsilon_{1,2}$. Using the definition of $g$ then gives Eq.~\eqref{eq:SM_applications_28}. However, as explained in section~\ref{sec:HMF}, this is only valid at sufficiently high temperature.

\subsection{Single impurity Anderson model}
\label{sec:SM-SIAM}

We now consider an interacting system, the single impurity Anderson model (SIAM), with Hamiltonian 
\begin{equation}
H_S=\varepsilon_d\,(n_\uparrow+n_\downarrow)+U\,n_\uparrow n_\downarrow,
\qquad n_\sigma=d_\sigma^{\dagger}d_\sigma,\label{eq:SM_applications_31}
\end{equation}
coupled to a spinful reservoir,
\begin{equation}
H=H_S+\sum_{\sigma}\int\! d\varepsilon\,\varepsilon\,a^{\dagger}_{\varepsilon\sigma}a_{\varepsilon\sigma}
+\lambda\sum_{\sigma}\int\! d\varepsilon\,\sqrt{J(\varepsilon)}\big(d_\sigma^{\dagger}a_{\varepsilon\sigma}+a^{\dagger}_{\varepsilon\sigma}d_\sigma\big).
\label{eq:SM_applications_32}
\end{equation}

\vspace{0.5cm}

The system has four Fock states which are all eigenstates of the grand-canonical Hamiltonian
\begin{equation}
\begin{aligned}
&\ket{0}, && \ket{\sigma}=d_\sigma^{\dagger}\ket{0}, && \ket{2}=d_\uparrow^{\dagger}d_\downarrow^{\dagger}\ket{0},\\
&K_S\ket{0}=0, && K_S\ket{\sigma}=(\varepsilon_d-\mu)\ket{\sigma}, && K_S\ket{2}=\big(2(\varepsilon_d-\mu)+U\big)\ket{2}.
\end{aligned}\label{eq:SM_applications_33}
\end{equation}
Unlike in the previous two examples, $d_\sigma$ is not an eigenoperator. Removing a particle of spin $\sigma$ changes the energy by an amount that depends on whether $\bar\sigma$ is occupied. The two Bohr frequencies, corresponding to adding a particle to the empty and to the singly occupied level, are
\begin{equation}
\omega_1\equiv\mu-\varepsilon_d,
\qquad
\omega_2\equiv\mu-\varepsilon_d-U .\label{eq:SM_applications_34}
\end{equation}
Since $[K_S,d_\sigma(1-n_{\bar\sigma})]=\omega_1\,d_\sigma(1-n_{\bar\sigma})$ and $[K_S,d_\sigma n_{\bar\sigma}]=\omega_2\,d_\sigma n_{\bar\sigma}$, each channel splits into two Bohr-frequency components,
$X_{\sigma,\omega_1}=d_\sigma(1-n_{\bar\sigma})$,  $X_{\sigma,\omega_2}=d_\sigma n_{\bar\sigma}$, so that $X_\sigma=X_{\sigma,\omega_1}+X_{\sigma,\omega_2}$. The products entering the diagonal part are projectors onto Fock states,
\begin{equation}
\begin{aligned}
X_{\sigma,\omega_1}^{\dagger}X_{\sigma,\omega_1}&=\ket{\sigma}\bra{\sigma}, &\qquad X_{\sigma,\omega_1}X_{\sigma,\omega_1}^{\dagger}&=\ket{0}\bra{0},\\
X_{\sigma,\omega_2}^{\dagger}X_{\sigma,\omega_2}&=\ket{2}\bra{2}, &\qquad X_{\sigma,\omega_2}X_{\sigma,\omega_2}^{\dagger}&=\ket{\bar\sigma}\bra{\bar\sigma}.
\end{aligned}
\label{eq:SM_applications_36}
\end{equation}
Writing $P_k=\ket{k}\bra{k}$ for $k\in\{0,\uparrow,\downarrow,2\}$ and $p^{(0)}_k=\langle P_k\rangle_{\tau_S}$, the diagonal part is
\begin{align}
\Delta_{\rm diag}=\sum_\sigma\Big[{}&\tau_S\big(P_\sigma-p^{(0)}_\sigma\big)D_1(\omega_1)+\tau_S\big(P_0-p^{(0)}_0\big)D_2(\omega_1)\nonumber\\
&+\tau_S\big(P_2-p^{(0)}_2\big)D_1(\omega_2)+\tau_S\big(P_{\bar\sigma}-p^{(0)}_{\bar\sigma}\big)D_2(\omega_2)\Big].\label{eq:SM_applications_37}
\end{align}
The off-diagonal part pairs $X_{\sigma,\omega_1}$ with $X_{\sigma,\omega_2}$. Every such product contains the factor $n_{\bar\sigma}(1-n_{\bar\sigma})=0$, meaning $\Delta_{\rm off}=0$. The state $\tau_{\MF}^{(2)}$ is therefore diagonal in the Fock basis as required by conservation of charge and $S_z$.

The bare Gibbs state is $\tau_S=\sum_k p^{(0)}_k P_k$ with
\begin{equation}
p^{(0)}_0=\frac{1}{Z_S},\quad p^{(0)}_\sigma=\frac{e^{\beta\omega_1}}{Z_S},\quad p^{(0)}_2=\frac{e^{\beta(\omega_1+\omega_2)}}{Z_S},\quad Z_S=1+2e^{\beta\omega_1}+e^{\beta(\omega_1+\omega_2)} .\label{eq:SM_applications_38}
\end{equation}
Since $P_k$ and $\tau_S$ commute, $\Trs\big[P_k\,\tau_S(P_{k'}-p^{(0)}_{k'})\big]=p^{(0)}_k\big(\delta_{kk'}-p^{(0)}_{k'}\big)$. Each state therefore collects $D_1$ from every transition in which it is in the state with one more particle, and $D_2$ from every transition in which it is in the state with one less particle. Defining
\begin{equation}
w_0=2D_2(\omega_1),
\quad
w_\sigma=D_1(\omega_1)+D_2(\omega_2),
\quad
w_2=2D_1(\omega_2),
\quad
\bar w\equiv\sum_k p^{(0)}_k w_k,
\label{eq:SM_applications_39}
\end{equation}
we obtain
\begin{equation}
\tau_{\MF}^{(2)}=\sum_{k}p_k^{(0)}\Big[1+\lambda^2\big(w_k-\bar w\big)\Big]\ket{k}\bra{k}.
\label{eq:SM_applications_40}
\end{equation}
The corrections to the double occupancy and to the spin-resolved occupation follow directly,
\begin{equation}
\delta\langle n_\uparrow n_\downarrow\rangle=\lambda^2p_2^{(0)}\big(w_2-\bar w\big),
\qquad
\delta\langle n_\sigma\rangle=\lambda^2\Big[p_\sigma^{(0)}\big(w_\sigma-\bar w\big)+p_2^{(0)}\big(w_2-\bar w\big)\Big],
\label{eq:SM_applications_41}
\end{equation}
The Hamiltonian of mean force is a renormalised SIAM, 
\begin{equation}
  H_{\MF}=\tilde\varepsilon_d\,(n_\uparrow+n_\downarrow)
    +\tilde U\,n_\uparrow n_\downarrow-\frac{\lambda^2}{\beta}w_0,
  \label{eq:SM_applications_42}
\end{equation}
where
\begin{align}
  \tilde\varepsilon_d-\varepsilon_d&=-\frac{\lambda^2}{\beta}\big(w_\sigma-w_0\big)
    =-\frac{\lambda^2}{\beta}\big[D_1(\omega_1)+D_2(\omega_2)-2D_2(\omega_1)\big],\nonumber\\
  \tilde U-U&=-\frac{\lambda^2}{\beta}\big(w_0-2w_\sigma+w_2\big)
    =-\frac{2\lambda^2}{\beta}\Big\{\big[D_1(\omega_2)-D_2(\omega_2)\big]-\big[D_1(\omega_1)-D_2(\omega_1)\big]\Big\}.\label{eq:SM_applications_43}
\end{align}
\section{Non-interacting fermions} \label{sec:sm-noninteracting}
If $H$ is quadratic in mode operators and particle number conserving, then the exact dynamics can be obtained via unitary evolution of the single-particle correlation matrix $\boldsymbol{C}_{ij}(t) = \textrm{Tr}(\rho(t)d^{\dag}_{j}d_{i})$ over all the system and reservoir modes  $\{d_i\} = (s_1,\dots,s_{L_S},a_1,\dots,a_{L_R})$. The ordering of the indices here is chosen such that the matrix $\boldsymbol{C}$ evolves unitarily, as we will now show. Defining $U(t) = e^{-iHt}$, we have
\begin{align} 
\boldsymbol{C}_{ij}(t) &= \bra{\psi(t)}d^{\dag}_{j}d_{i}\ket{\psi(t)} \nonumber = \bra{\psi(0)}U^{\dag}(t)d^{\dag}_{j}U(t)U^{\dag}(t)d_{i}U(t)\ket{\psi(0)} \nonumber \\
&= \bra{\psi(0)}d^{\dag}_{j}(t)d_{i}(t)\ket{\psi(0)}.\label{eq:SM_noninteracting_1}
\end{align}
Differentiating $d^{\dag}_{j}(t)$ gives 
\begin{equation} \label{eq:SM_noninteracting_2}
\frac{\textrm{d}d^{\dag}_{j}(t)}{\textrm{d}t} = i U^{\dag}(t)[H,d_{j}^{\dag}]U(t) = i\sum_{k}\boldsymbol{H}_{kj}d_{k}^{\dag}(t)
\end{equation}
where we have used $[H,d_{j}^{\dag}] = \sum_{k}\boldsymbol{H}_{kj}d_{k}^{\dag}$ for $H=\sum_{kl}\boldsymbol{H}_{kl}d_{k}^\dag d_{l}$. Defining the vector of operators $\underline{d}^{\dag}(t) = (d_{1}^{\dag}(t),...,d_{L}^{\dag}(t))^{T}$, where $L=L_{S}+L_{R}$, we have 
\begin{equation} \label{eq:SM_noninteracting_3}
\frac{\textrm{d}\underline{d}^{\dag}(t)}{\textrm{d}t} = i\boldsymbol{H}^{T}\underline{d}^{\dag}(t).
\end{equation} 
which integrates to $\underline{d}^{\dag}(t) = \boldsymbol{U}^{*}(t)\underline{d}^{\dag}(0)$ with $\boldsymbol{U}(t) \equiv e^{-i\boldsymbol{H}t}$. Substituting this into Eq.~\eqref{eq:SM_noninteracting_1} gives the result
\begin{align}  
\boldsymbol{C}_{ij}(t) &= \bra{\psi(0)}\sum_{kl} \boldsymbol{U}^\dagger_{kj}d^\dagger_{k}(\boldsymbol{U}^\dagger_{li}d^\dagger_{l})^\dagger \ket{\psi(0)} = \sum_{kl} \boldsymbol{U}^\dagger_{kj}\boldsymbol{U}_{il}\bra{\psi(0)}d^{\dagger}_{k}d_{l}\ket{\psi(0)} \nonumber \\
&= \sum_{kl} \boldsymbol{U}^\dagger_{kj}\boldsymbol{U}_{il} \boldsymbol{C}_{lk}(0).\label{eq:SM_noninteracting_4}
\end{align}
Thus we have
\begin{equation}  \label{eq:SM_noninteracting_5}
\boldsymbol{C}(t) = \boldsymbol{U}\boldsymbol{C}(0)\boldsymbol{U}^{\dag}.
\end{equation}

\subsection{Exact mean force Gibbs state from the correlation matrix}
\label{sec:SM-static-C}

For quadratic models the correlation matrix also gives the equilibrium state directly, without any time evolution. For simplicity, we consider a finite number of modes, but the continuum can be taken straightforwardly. We write the generator $K$ of Eq.~\eqref{eq:SM_derivation_5} in terms of a single particle matrix $\boldsymbol{K}_{ij}$ and diagonalise it in its eigenmodes,
\begin{equation}
K=\sum_{ij}\boldsymbol{K}_{ij}\,d_i^{\dagger}d_j = \sum_n w_nb_n^{\dag}b_{n}.\label{eq:SM_noninteracting_6}
\end{equation}
The global Gibbs state factorises over the normal modes $\{b_n\}$, with $\langle b_{n}^{\dag}b_{m}\rangle = \delta_{nm}f(w_n)$, where $f(w)$ is the Fermi function. In the original basis, the single particle correlation matrix of the global Gibbs state is
\begin{equation}
\boldsymbol{C}_{ij}\equiv\Trs\big(\tau_{\rm tot}\, d_j^{\dagger}d_i\big)=\big[f(\boldsymbol{K})\big]_{ij}.\label{eq:SM_noninteracting_7}
\end{equation}
Within the system block, we have $\boldsymbol{C}_S = [f(\boldsymbol{K})]_{i,j\in S}$. For quadratic fermions, the reduced density matrix of a subsystem (provided the state is parity symmetric) is given in terms of the correlation matrix via~\cite{Peschel2003,PhysRevB.69.075111,Strachan2024}
\begin{equation}
\tau_{\MF}=\det\!\big(\mathbb{1}-\boldsymbol{C}_S\big)\,
\exp\!\Big\{\sum_{ij\in S}\Big[\ln\!\big(\boldsymbol{C}_S(\mathbb{1}-\boldsymbol{C}_S)^{-1}\big)\Big]_{ij}\,s_i^{\dagger}s_j\Big\},\label{eq:SM_noninteracting_8}
\end{equation}
or equivalently
\begin{equation}
\tau_{\MF}=\frac{\exp\!\big(-\sum_{ij}\boldsymbol{M}_{ij}\,s_i^{\dagger}s_j\big)}{\Trs_S\exp\!\big(-\sum_{ij}\boldsymbol{M}_{ij}\,s_i^{\dagger}s_j\big)},
\qquad
\boldsymbol{M}=\ln\!\big(\boldsymbol{C}_S^{-1}-\mathbb{1}\big).\label{eq:SM_noninteracting_9}
\end{equation}
$\beta^{-1}\boldsymbol{M}$ is thus the single-particle matrix of $K_\textrm{MF} = H_\textrm{MF}-\mu N_S$ with temperature dependence through $\boldsymbol{C}_S$.

\end{document}